\documentclass[twocolumn]{aastex631}

\newcommand\gaia{{\em Gaia\ }}
\newcommand\rguide{$R_{guide}$~}
\newcommand\rgc{$R_{GC}~$}
\newcommand\occam{OCCAM~}
\usepackage{amsmath}
\usepackage{moresize}
\usepackage{soul}
\accepted{\today}

\submitjournal{AJ}

\shorttitle{OCCAM VI.: Abundance Gradients from APOGEE DR17}
\shortauthors{Myers et al.}

\begin{document}

\title{The Open Cluster Chemical Abundances and Mapping Survey: \\ VI. Galactic Chemical Gradient Analysis from APOGEE DR17}

\correspondingauthor{Natalie Myers}
\email{n.myers@tcu.edu}

\author[0000-0001-9738-4829]{Natalie Myers}
\affiliation{Department of Physics and Astronomy, Texas Christian University, TCU Box 298840 \\
Fort Worth, TX 76129, USA }

\author{John Donor}
\affiliation{Department of Physics and Astronomy, Texas Christian University, TCU Box 298840 \\
    Fort Worth, TX 76129, USA }

\author[0000-0003-4019-5167]{Taylor Spoo}
\affiliation{Department of Physics and Astronomy, Texas Christian University, TCU Box 298840 \\
    Fort Worth, TX 76129, USA }

\author[0000-0002-0740-8346]{Peter M.~Frinchaboy}
\affiliation{Department of Physics and Astronomy, Texas Christian University, TCU Box 298840 \\
Fort Worth, TX 76129, USA }

\author{Katia Cunha}
\affiliation{Observatório Nacional, Rua General José Cristino, 77, Rio de Janeiro, RJ 20921-400, Brazil}
\affiliation{Steward Observatory, University of Arizona, 933 North Cherry Avenue, Tucson, AZ 85721-0065, USA}

\author[0000-0003-0872-7098]{Adrian M.~Price-Whelan}
\affiliation{Center for Computational Astrophysics, Flatiron Institute, 162 Fifth Avenue, New York, NY 10010, USA}

\author{Steven R. Majewski}
\affil{Department of Astronomy, University of Virginia, Charlottesville, VA 22904-4325, USA}

\author[0000-0002-1691-8217]{Rachael~L.~Beaton}
\affiliation{Department of Astrophysical Sciences, Princeton University, 4 Ivy Lane, Princeton, NJ~08544}
\affiliation{The Observatories of the Carnegie Institution for Science, 813 Santa Barbara St., Pasadena, CA~91101}

\author[0000-0001-6761-9359]{Gail Zasowski}
 \affiliation{Department of Physics \& Astronomy, University of Utah, 115 S. 1400 E., Salt Lake City, UT 84112, USA}

\author[0000-0003-2321-950X]{Julia O'Connell}
\affiliation{Department of Physics and Astronomy, Texas Christian University, TCU Box 298840 \\
    Fort Worth, TX 76129, USA }

\author{Amy E. Ray}
\affiliation{Department of Physics and Astronomy, Texas Christian University, TCU Box 298840 \\
Fort Worth, TX 76129, USA }

\author{Dmitry Bizyaev}
\affiliation{Apache Point Observatory and New Mexico State
University, P.O. Box 59, Sunspot, NM, 88349-0059, USA}
\affiliation{Sternberg Astronomical Institute, Moscow State
University, Moscow, Russia}

\author{Cristina Chiappini}
\affil{Leibniz-Institut f\"ur  Astrophysik Potsdam (AIP), An der Sternwarte 16, 14482 Potsdam, Germany}

\author[0000-0002-1693-2721]{D. A. Garc{\'i}a-Hern{\'a}ndez}
\affiliation{Instituto de Astrof{\'i}sica de Canarias, V{\'i}a L\'actea S/N, 38205 La Laguna, Tenerife, Spain}
\affiliation{Universidad de La Laguna, Departamento de Astrof{\'i}sica, 30206 La Laguna, Tenerife, Spain}

\author{Doug Geisler}
\affiliation{Departamento de Astronom{\'i}a, Universidad de Concepci{\'o}n, Casilla 160-C, Concepci{\'o}n, Chile}
\affiliation{Instituto de Investigaci{\'o}n Multidisciplinario en Ciencia y Tecnolog{\'i}a, Universidad de La
Serena, \\Avenida Ra{\'u}l Bitr{\'a}n S/N, La Serena, Chile}
\affiliation{Department of Astronomy, Universidad La Serena, La Serena, Chile}

\author[0000-0002-4912-8609]{Henrik J\"onsson}
\affil{Materials Science and Applied Mathematics, Malm\"o University, SE-205 06 Malm\"o, Sweden}

\author[0000-0003-1805-0316]{Richard R. Lane}
\affiliation{Centro de Investigaci{\'o}n en Astronom{\'ia}, Universidad Bernardo O'Higgins, Avenida Viel 1497, Santiago, Chile}

\author{Pen{\'e}lope Longa-Pe{\~n}a}
\affiliation{Centro de Astronom{\'i}a (CITEVA), Universidad de Antofagasta, Avenida Angamos 601, Antofagasta 1270300, Chile}

\author[0000-0002-5627-0355]{Ivan Minchev}
\affiliation{Leibniz-Institut f\"ur  Astrophysik Potsdam (AIP), An der Sternwarte 16, 14482 Potsdam, Germany}

\author[0000-0002-7064-099X]{Dante Minniti}
\affiliation{Departamento de Ciencias Fisicas, Facultad de Ciencias Exactas, Universidad Andres Bello, Av. \\Fernandez Concha 700, Las Condes, Santiago, Chile}
\affiliation{Vatican Observatory, V00120 Vatican City State, Italy}

\author[0000-0003-4752-4365]{Christian Nitschelm}
\affiliation{Centro de Astronom{\'i}a (CITEVA), Universidad de Antofagasta, Avenida Angamos 601, Antofagasta 1270300, Chile}

\author[0000-0002-1379-4204]{A. Roman-Lopes} 
\affiliation{Department of Astronomy, Universidad La Serena, La Serena, Chile}

\begin{abstract}
The goal of the Open Cluster Chemical Abundances and Mapping (OCCAM) survey is to constrain key Galactic dynamic and chemical evolution parameters by the construction and analysis of a large, comprehensive, uniform data set of infrared spectra for stars in hundreds of open clusters. 
This sixth contribution from the OCCAM survey presents analysis of SDSS/APOGEE Data Release 17 (DR17) results for a sample of stars in 150 open clusters, 94 of which we designate to be ``high quality'' based on the appearance of their color-magnitude diagram. 
We find the APOGEE DR17-derived [Fe/H] values to be in good agreement with those from previous high resolution spectroscopic open cluster abundance studies. 
Using {a subset of} the high quality sample, the Galactic abundance gradients were measured for 16 chemical elements, including [Fe/H], for both {Galactocentric radius} ($R_{GC}$) and {guiding center radius} ($R_{Guide}$). We find an overall Galactic [Fe/H] vs $R_{GC}$ gradient of $-0.073 \pm 0.002$ dex/kpc over the range of $6 <$ \rgc $< 11.5$ kpc, and a similar gradient is found for [Fe/H] versus $R_{Guide}$. 
Significant Galactic abundance gradients are also noted for O, Mg, S, Ca, Mn, Na, Al, K and Ce. Our large sample additionally allows us to explore the evolution of the gradients in four age bins for {the remaining} 15 elements.

\end{abstract}


\section{Introduction} \label{sec:intro}

Open clusters are key, age-datable tracers that have long been used to explore chemical trends in the Galactic disk. Since the early work of \citet{janes_79}, numerous studies have advanced the field, particularly over the past 15 years 
\citep[e.g.,][]{sestito2008, bragaglia2008, friel2010, carrera_2011, yong_2012, frinchaboy_13, reddy_16, occam_katia,  netopil_16, magrini_2017,  occam_p4, spina21, Netopil21}, with
progress driven by the availability of larger telescopes, the expansion of multi-fiber spectroscopic capabilities, and, more recently, by large-scale high-resolution spectroscopic surveys, such as, Gaia-ESO \citep{gaia_eso}, GALactic Archeology with HERMES \citep[GALAH;][]{galah} and Apache Point Observatory Galactic Evolution Experiment \citep[APOGEE;][]{apogee}. 
The APOGEE-based Open Cluster Chemical Abundances and Mapping  (OCCAM) Survey has produced a comprehensive, uniformly measured data set of infrared spectra for stars in over a hundred open clusters, with the goal of exploiting the advantages of open clusters for constraining key Galactic dynamic and chemical parameters. 

Since our previous open cluster chemical abundance gradient study \citep[][OCCAM-IV, hereafter]{occam_p4}, which was based on SDSS-IV APOGEE Data Release 16 (DR16; \citet{dr16}), a few new studies of Galactic abundance gradients have been published, most having incorporated the data from OCCAM-IV. 
{For example, \citet{Zhang21} use a compilation of LAMOST, APOGEE, and other surveys to constrain the metallicity gradient and acquire a high quality sample of young open clusters. \citet{Netopil21} also use APOGEE clusters and other studies to  characterize the metallicity gradient and its evolution over eight different age bins.} Both studies explore the potential effects of radial migration in open clusters, which is possible due to the availability of high quality kinematic data from the ESA \textit{Gaia} mission \citep{gaia_mission}. The results from the ESA \textit{Gaia} mission have also significantly improved the ability to refine cluster membership, which is utilized in {many studies \citep[e.g.,][]{cg_18, cg20, castroginard_21, Kounkel_20, Monteiro_19, dias_21}}.

Galactic abundance gradients are important observable constraints to models of Milky Way chemical evolution, but limitations to these constraints arise from {(1)} the use of inhomogenous or small datasets, {(2)} systematic offsets in the abundance results when combining data sets, and {(3)} uncertainties in cluster ages and distances adopted in the different studies. For example, \citet{donor_18} found a 40\% variation in the gradient slopes when using different distance catalogs but the same set of abundance results. Complications like these have led to a range of values for the metallicity gradients derived from open cluster samples --- between roughly $-0.05 ${$\pm 0.011$} dex/kpc \citep{reddy_16, casamiquela_2019} and $-0.1 \pm 0.02$ dex/kpc \citep{jacobson_16}. 

Even more recent studies have been able to break down the iron gradient into individual elements to further investigate the processes which enrich the Milky Way, for instance, 
OCCAM-IV, \citet{spina21,spina22},
and \citet{apogee_Ce}, all do so with large open cluster samples.

In this paper, we present the complete OCCAM sample, which is based on the APOGEE results given in the SDSS-IV Data Release 17 (DR17) \citep{sdss4,dr17}, the most recent and final release of data products from Apache Point Observatory Galactic Evolution Experiment 2 \citep[APOGEE-2;][Holtzman et al., \textit{in prep}]{apogee}. We analyze Galactic gradient trends in metallicity ([Fe/H]), $\alpha$ elements (O, Mg, Si, S, Ca, Ti), iron-peak elements (V, Cr, Mn, Co, Ni), and other elements (Na, Al, K, Ce) represented in the APOGEE DR17 database, and explore the evolution of these gradients as a function of age.  We also calculate the trends with Galactocentric guiding center radius ($R_{guide}$) to investigate the potential biases that may affect the analysis by using the current cluster locations. Finally, we discuss this sample in comparison to other recent literature studies of open clusters.

\begin{figure*}[t!]
    \epsscale{1.2}
    \centering
    \plotone{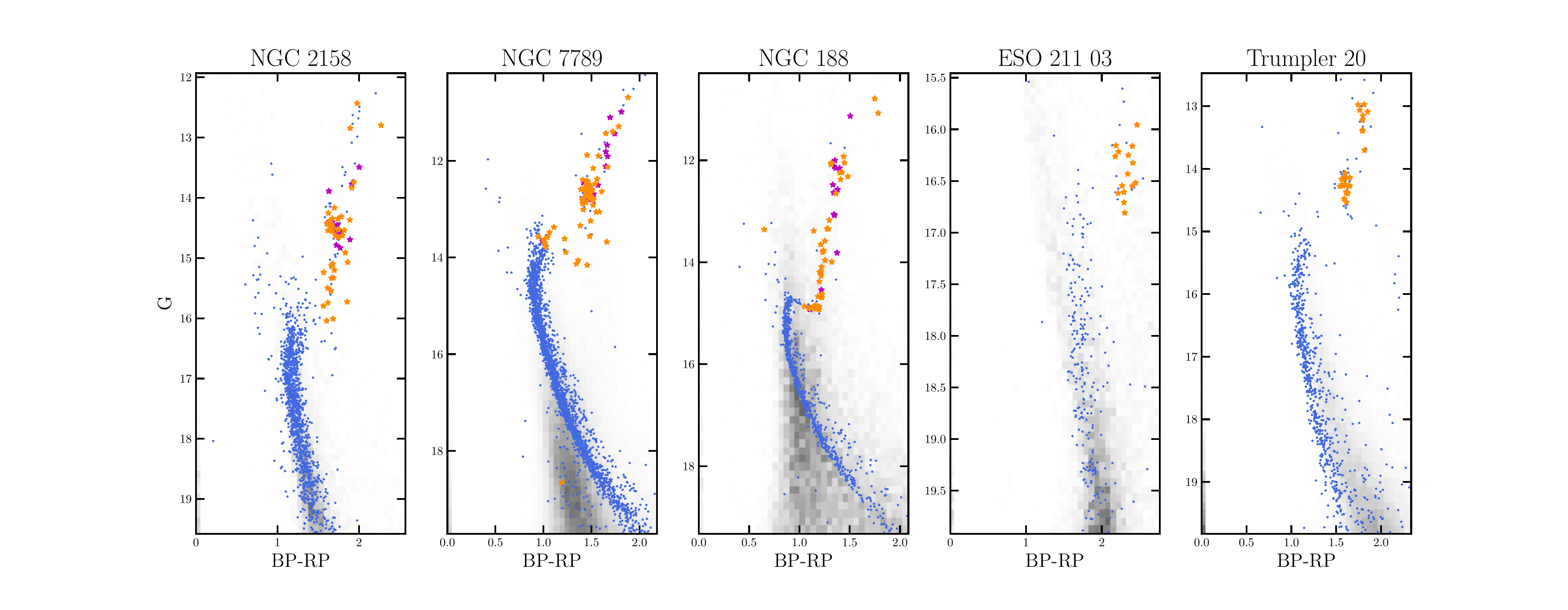} 
 	\caption{ \small Five example color-magnitude diagrams of open clusters analyzed in this study (Table \ref{tab:full_sample}). Stars from \gaia EDR3 {within twice the cluster radius, as defined in \citealt{cg20}, are {included}; 
 	 stars} identified as PM members {and inside the cluster radius} are  {blue}.  {Non-member stars are shown as a Hess diagram in grey}. 
 	The OCCAM pipeline-identified APOGEE members from DR16 \citep{occam_p4} are shown in purple.  New DR17 OCCAM pipeline-identified APOGEE member stars are shown as orange stars.
 	}
 	\label{fig:qual}
 \end{figure*}

\section{Data} \label{sec:data}
To minimize the systemic offsets inherent to blending multiple data sources, we create a uniform sample for our abundance gradient analyses by pulling the majority of our data from only two sources: SDSS/APOGEE and \textit{Gaia}. 
To supplement the SDSS/APOGEE data and provide astrometric and photometric parameters for our analysis we use data from $3,720,692$ \gaia {Early Data Release 3 \citep[EDR3][]{gaia_edr3}} stars and radial velocities for $38,667$ \gaia stars \citep{gaia_edr3_rvs}. We check the offset between Gaia EDR3 RVs and APOGEE DR17 RVs and find the median offset to be roughly $-0.14$ km/s with a standard deviation of approximately 2.80 km/s.

\subsection{SDSS/APOGEE DR17}

The chemical abundances and radial velocities for the open cluster stars in our sample are from the APOGEE SDSS-IV values in DR17 
\citep[][]{sdss4,dr17}.  As previously mentioned, this final data release derives data from the completed collection of 
high resolution, near-infrared spectra taken with the APOGEE spectrographs \citep{spectrograph} for over 650,000 stars as part of the APOGEE and APOGEE-2 surveys.  The SDSS/APOGEE data were taken using two telescopes: the Sloan Foundation telescope at the Apache Point Observatory \citep[New Mexico, APO][]{sloan_telescope} in the Northern Hemisphere and Du Pont telescope at the Las Campanas Observatory \citep[Chile, LCO][]{du_pont} in the Southern Hemisphere. 
Observations for APOGEE-2N were concluded in November 2020, while those for APOGEE-2S were concluded in January 2021.

Targeting for the APOGEE survey, including details from this program, are described in \citet{AP_Target_selection}, \citet{zasowski13}, \citet{zasowski17}, \citet{ap2n_target}, and \citet{ap2s_target}.  APOGEE data are reduced using the APOGEE data reduction pipeline \citep{nidever_2015} and the ASPCAP data analysis pipeline \citep{aspcap}. The latter produces the detailed abundances of chemical elements  \citep[][Holtzman et al., {\it in prep}]{holtzman_2015,holtzman_2018,jonsson_2020} that are central to the present study.

\subsubsection{Pipeline Changes in APOGEE DR17}

Some significant changes were made to the APOGEE pipeline for DR17. New synthetic spectral libraries were created using the Synspec code \citep{synspec17,synspec21} that now allows for the Non-LTE analysis of the elements  Na, Mg, K, and Ca using the computations in \citet{nlte_calcs}.
The APOGEE line list used for DR17 is \citet{smith21}, which updates from \citet{shetrone_2015}. While schematic description of the DR17 pipeline is given in \citet{dr17}, further specifics about updates to the APOGEE pipeline will be discussed in Holtzman et al. (\textit{in prep}).

\section{Methods} \label{sec:methods}
To identify cluster member stars, we employ the analysis described in \citet{donor_18, occam_p4}, which uses the celestial coordinates (RA/Dec), proper motion (PM), radial velocity (RV), and metallicity of stellar candidates to sift likely cluster members from non-members. All stars designated as cluster members must have RV, PM, and [Fe/H] values within 3$\sigma$ of the cluster mean. For a more thorough discussion of the probability values, see \S \ref{sec:vac}.
As in OCCAM-IV, we also use visual quality checks of both the color magnitude diagrams (CMDs) and Kiel diagrams ($T_{eff}$ vs $log(g)$) for the APOGEE stars in each cluster to distinguish between high quality clusters (with quality flag 1 {and 2\footnote{{The quality flag of 2 denotes a cluster used in the calibration sample from \citet{donor_18}.}}}) and potentially unreliable clusters (with quality flag 0). This procedure is discussed in more detail in OCCAM-IV.
As an example of both the CMDs used and the difference between APOGEE DR16 and DR17, we show five example clusters, (all with a quality flag of 1 or 2), in Figure \ref{fig:qual}. As can be seen, the addition of APOGEE DR17 not only expands the number of stars that are identified to be likely cluster members in previously known clusters, but it also allows new clusters to be added to our sample.

\subsection{Methodology Changes from \citet{occam_p4}}

The present analysis adopts several changes in methodology from that employed in \citet{occam_p4}.
In addition to using the latest stellar parameters and abundances from the greatly expanded APOGEE DR17 sample, we also use the latest data from the \cite{gaia_mission}, {EDR3}, to take advantage of the extended baseline and expanded astrometric catalog. 
Additionally, we use the open cluster parameters from \citet{cg20}, which exclusively uses \gaia DR2 to compile a catalog that
provides uniform measurements of age and distance (among other parameters) for {roughly} 2,000 open clusters. 
This includes all the open clusters used here for analysis of the Galactic chemical gradients.

Another change in methodology applied to our analysis is the addition of the guiding center radius, $R_{guide}$, which is now used along with galactocentric radius, $R_{GC}$, to compute the Galactic abundance gradients. Methods for the calculation of \rguide are further discussed in \S \ref{sec:Rguide}.
\footnote{For two clusters (FSR 0542 and NGC 2232) that were not initially recovered using the OCCAM-IV pipeline, we implemented a parallax cut for stars greater than twice the reported distance to the cluster \citep{cg20} and those less than half the distance to the cluster.}

Finally, because more Ce II lines were used in DR17 to determine the abundance of Ce in ASPCAP, the cerium abundance measurements have significantly improved over those in DR16; as a result, we are able to report more precise Galactic trends in cerium here.

\subsection{Computing Guiding Center Radii \rguide} \label{sec:Rguide}

For each cluster in the sample, we compute its guiding-center radius \rguide using the circular velocity rotation curve from the best-fitting Milky Way model described in \citet{Price-Whelan:2021}.
The guiding center radius of a given 
general, eccentric orbit is the radius of a circular orbit with the same angular momentum as the generic orbit.
We compute the approximate guiding center radii for the OCCAM clusters by first transforming their heliocentric position and velocity data (sky position, distance, proper motions, and radial velocity) into Galactocentric Cartesian coordinates, assuming solar parameters: for the Sun--Galactic center distance we adopt $R_\odot = 8.275~\textrm{kpc}$ \citep{GRAVITY:2021}, a solar height above the Galactic midplane of $z_\odot = 20.8~\textrm{pc}$ \citep{Bennett:2019}, and a solar velocity with respect to the Galactic center $\boldsymbol{v}_\odot = (8.42, 250.2, 7.90)~\textrm{km}~\textrm{s}^{-1}$ \citep{Drimmel:2018, GRAVITY:2018, Reid:2004}. 
We then compute the $z$-component of the angular momentum vector $L_z$ for each cluster in the Galactocentric frame and estimate the guiding center radii as $R_{guide} = L_z / v_c(R)$, where $R$ is the present-day cylindrical radius of each cluster and $v_c(r)$ is the circular velocity curve evaluated at the radius of each cluster.

{The use of the guiding center radius of a cluster, rather than its present galactocentric radius, has the advantage of correcting for orbital blurring effects in the metallicity gradients (e.g., \citealt{Netopil21}, \citealt{Zhang21}, \citealt{spina21}). 
To illustrate and explore the differences between \rguide and $R_{GC}$, we calculate $R_{guide}$, and discuss both radii in \S \ref{sec:results} and \S \ref{sec:discussion}.}

\subsection{Membership Comparison to \citet{cg_18}}

\begin{figure*}[t]
    \begin{center}
    \epsscale{1.2}
 	\plotone{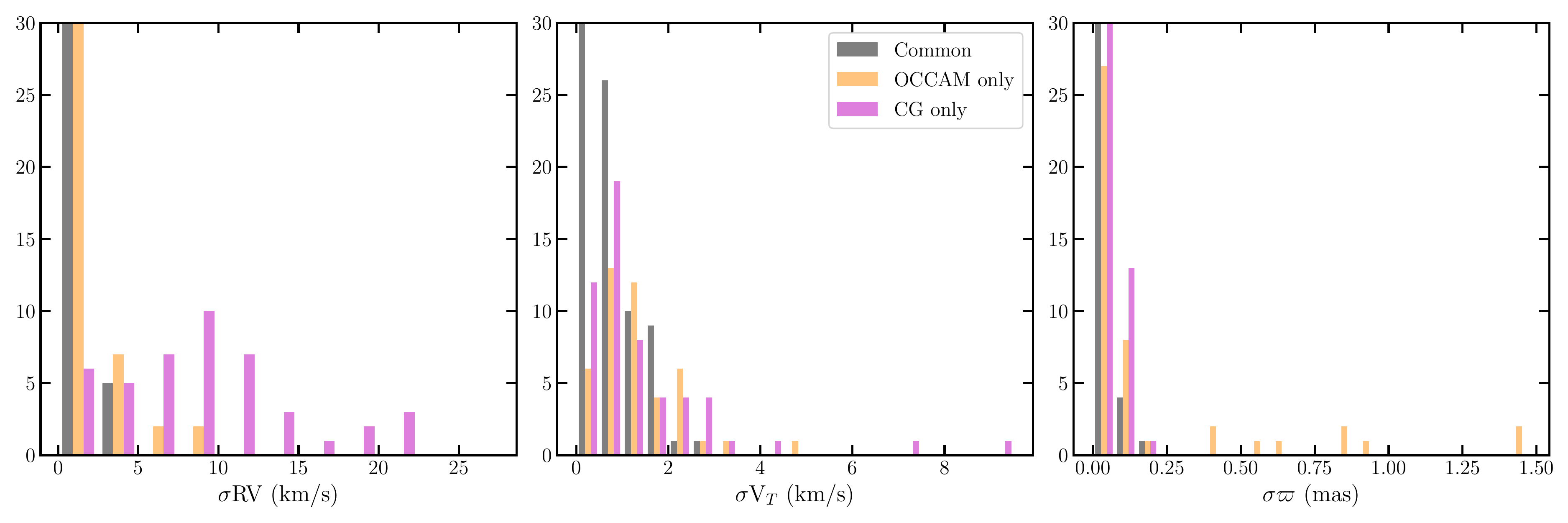}
 	\vskip0.1in
 	\caption{{\small We present the measured standard deviations of three key cluster properties (radial velocity, transverse velocity, and parallax) using three different membership sub-samples: 1) includes member stars common to both our \occam analysis and \citet{cg_18}, 2) \occam members that are not \citet{cg_18} members (\occam only), and 3) \citet{cg_18} only member stars (CG only). NOTE: the histogram are artificially cut-off at 30 to show relevant detail.}}
 	\label{fig:CGcompare}
 	\end{center}
\end{figure*}

{
\citet{cg_18} performed a fundamentally different membership analysis than presented here and previously by OCCAM \citep{donor_18,occam_p4}. Whereas our analysis relies on kernel convolution and Gaussian fitting to define a rigid boundary for what constitutes ``the cluster'', \citet{cg_18} performed a clustering search in the 5 dimensional \gaia phase space (RA, dec, proper motion, and parallax [$\varpi$]), requiring no fitting or boundary setting. Besides the difference in methodology, the absence of a constraint on parallax in our method, and inclusion of RV and [Fe/H], are noteworthy.}

{In order to compare the results from the two methods, we divide stars into three categories: common (stars considered cluster members in both the OCCAM and the CG samples), OCCAM only (stars that are included in the present sample but not in \citealt{cg_18}), and CG only (stars rejected from the present sample but included in \citealt{cg_18}). We create a statistic that accentuates differences between these three samples in RV, transverse velocity ($V_T$, calculated using the cluster distance measured by \citealt{cg20}), and parallax ($\varpi$). To compute this statistic, we first measure the mean cluster value for stars in the common sample ($\bar{x}_{common}$). We then compute the average deviation of the OCCAM only and the  CG only samples ($x_{single \textrm{ } sample}$) from $\bar{x}_{common}$, shown in Eq. \ref{eq:modstd}}.
{\begin{equation}\label{eq:modstd}
    \sigma_{mod} = \frac{1}{n} \sum \big( \bar{x}_{common} - x_{single \textrm{ } sample} \big)
\end{equation}
}

{In Figure \ref{fig:CGcompare}, we plot a histogram of RV, $V_T$, and $\varpi$ with (1) the $1\mbox{-}\sigma$ standard deviation of the common sample within each cluster in gray, (2) $\sigma_{mod}$ for OCCAM only in orange, and (3) $\sigma_{mod}$ for CG only in purple. 
There are 92 clusters in our sample where the results of our membership analysis differ from \citet{cg_18}; we omit the remaining clusters from this analysis as it is designed to show differences. To show relevant detail we artificially cut off each histogram at 30; in all 3 panels the lowest bin is populated beyond what is shown. 
Since the common sample is more restrictive than either individual sample it is not surprising that we measure a small standard deviation for the common sample of stars in most clusters. We note the scale difference between the RV and $V_T$ histograms: the first 5 bins in the $V_T$ plot span 0--2.5 km/s, which is the size of the first bin in the RV plot.}

{
For the common sample, 67 of 92 clusters for which we measure RV dispersion have measured dispersions below 1 km/s, and in $V_T$ we find 71 of 92 clusters showing a dispersion below 1 km/s, in good agreement with typical cluster dispersions \citep[e.g.,][]{cg_anders_20} and despite not explicitly removing binary stars. 
For $\varpi$ the majority of the OCCAM only clusters show low dispersions, comparable to the common sample and CG only sample, with 83 of 92 clusters having dispersions $< 0.25$ mas, 
despite the fact we have not used $\varpi$ in our selection. For the remaining nine clusters, five are in our low quality sample. Of the four in our high quality sample two are very nearby, Melotte 22 (The Pleiades) and Ruprecht 147, so some dispersion in $\varpi$ is expected. The remaining two clusters, FSR 0496 and NGC 7789, each have one star with negative $\varpi$ reported, significantly affecting the measured dispersion.
}

{This analysis shows that despite different selection criteria, the reliability of our sample is comparable to \citet{cg_18}, where we can compare directly. We find inconsistencies in $\varpi$ in our sample, which is not surprising since $\varpi$ is not accounted for in our analysis. Similarly, we show there is significant RV variation in the \citet{cg_18} sample since RV was not accounted for in that analysis. A union of the two samples is straightforward to create using the VAC discussed in \S \ref{sec:vac}. Combining the 5 dimensional \gaia phase space with RV and [Fe/H] would produce a purer sample, but in the present work we have chosen to continue using the OCCAM IV membership selection pipeline for consistency.
}

\begin{figure}
  \epsscale{1.35}
 \plotone{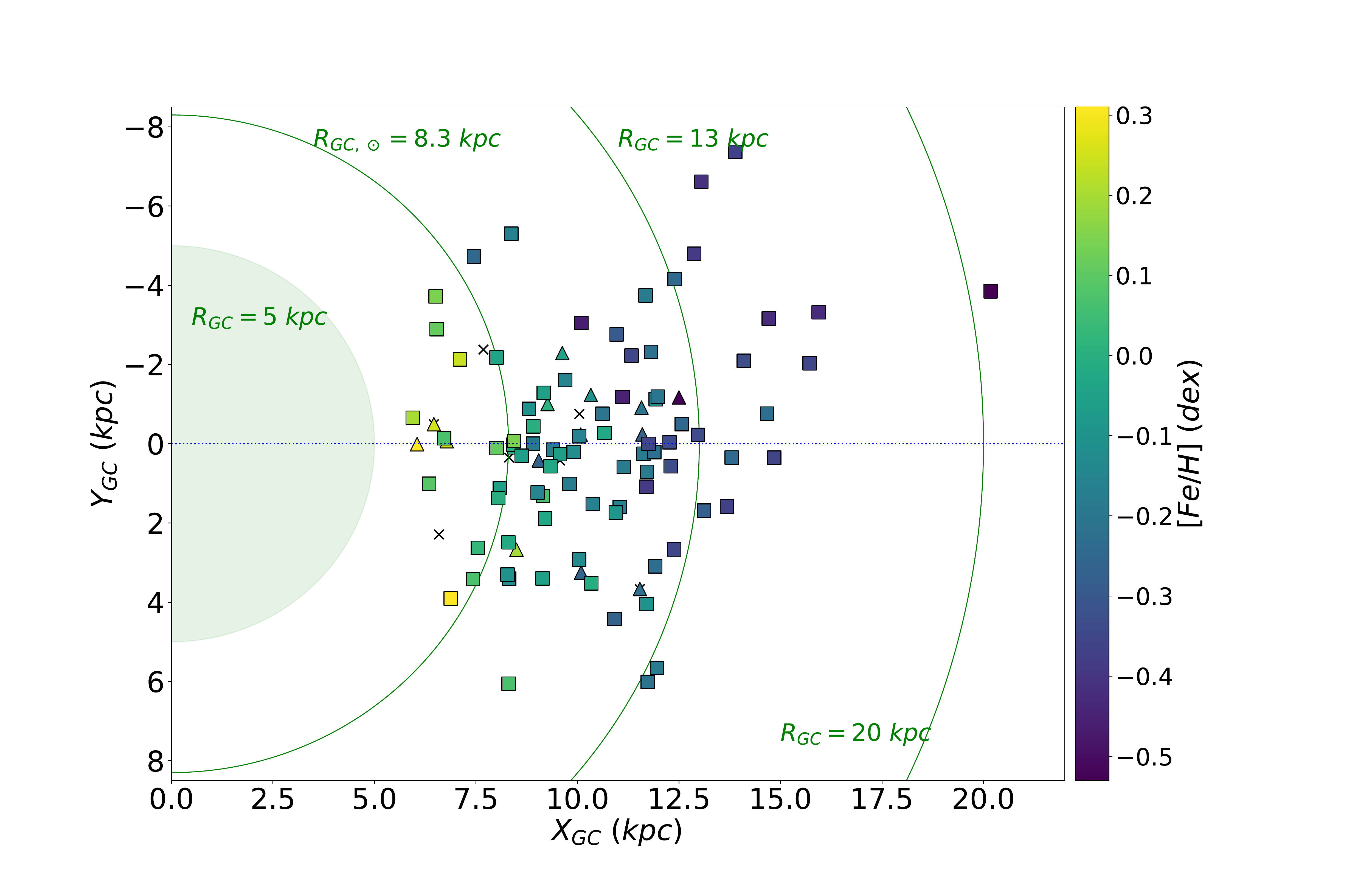}
 	\caption{ \small The OCCAM DR17 sample in common with \citet{cg20} plotted in the Galactic plane, color-coded by [Fe/H]. Square points are ``high quality'' clusters, triangles are the lower quality clusters, and crosses 
 	 {denote clusters which were in the ``high quality" sample of  OCCAM-IV but are now in the ``low quality" sample.}
    }
 	\label{fig:XYfull}
 \end{figure}

\section{The OCCAM DR17 Sample}\label{dr17}

Our final sample consists of 150 open clusters with 2061 member stars, out of $\sim26,700$ stars in the vicinity of a known open cluster
considered in this analysis. The final sample of clusters is shown in Figure \ref{fig:XYfull}. After a visual CMD inspection (described further in OCCAM-IV), we designate 94 clusters as ``high quality''. All clusters analyzed are presented in Table \ref{tab:full_sample} and Table \ref{tab:full_sample2}. Where Table \ref{tab:full_sample} includes bulk cluster parameters derived or adopted for this study, and Table \ref{tab:full_sample2} includes bulk cluster abundances, which are averaged over the stellar members.

For all Galactic abundance analysis in this study, we choose to use only clusters flagged as high quality and that have distances available from \citet{cg20}.
Additionally, we also cut out two clusters with an age less than 50 Myrs (NGC 7058 and Teutsch 1), due to previous studies suggesting the young star pipeline results from APOGEE may be unreliable \citep[e.g.,][]{kounkel_2018}. This results in a sample of 85 clusters.

\begin{deluxetable*}{lrrrrrrrrrrrrr}[b!]
\tabletypesize{\ssmall}
\tablecaption{OCCAM DR17 ``High Quality'' Sample - Basic Parameters \label{tab:full_sample}}
	\tablehead{
    \colhead{Cluster} &
    \colhead{Qual} &
    \colhead{l} &
    \colhead{b} &
    \colhead{R\tablenotemark{a}} &
     \colhead{Age\tablenotemark{a}} &
    \colhead{R$_{GC}$\tablenotemark{b}} & 
    \colhead{R$_{Guide}$\tablenotemark{b}} & 
    \colhead{$\mu_{\alpha}$\tablenotemark{c}} &
    \colhead{$\mu_{\delta}$\tablenotemark{c}} &
    \colhead{RV} & 
    \colhead{[Fe/H]} & 
    \colhead{Num}\\[-2ex] 
    \colhead{name} &
    \colhead{flag} &
    \colhead{deg} &
    \colhead{deg} &
    \colhead{(\arcmin)} &
     \colhead{Gyr} &
    \colhead{(kpc)} &
    \colhead{(kpc)} &
    \colhead{(mas yr$^{-1}$)} &
    \colhead{(mas yr$^{-1}$)} &
    \colhead{(km s$^{-1}$)} & 
    \colhead{(dex)} &
    \colhead{stars} 
    }
	\startdata
\multicolumn{11}{c}{High Quality Clusters}\\\hline
Berkeley 2   &  1 & 119.7032 & $ -2.3156$ &  3.2 & 0.59 & 12.92  & 13.16 & $-1.07 \pm 0.02$ & $-0.37 \pm 0.04$ & $-75.8 \pm 2.5$ & $-0.21 \pm 0.02$ &  6  \\
Berkeley 17  &  2 & 175.6578 & $ -3.6769$ &  8.5 & 7.24 & 11.33  & 11.60 & $+2.55 \pm 0.03$ & $-0.36 \pm 0.02$ & $-73.5 \pm 0.3$ & $-0.18 \pm 0.03$ &  8  \\
Berkeley 18  &  1 & 163.5891 & $  5.0296$ & 14.0 & 4.37 & 13.48  & 13.75 & $+0.75 \pm 0.03$ & $-0.09 \pm 0.02$ & $ -3.0 \pm 1.5$ & $-0.37 \pm 0.03$ & 30  \\
Berkeley 19  &  1 & 176.9168 & $ -3.6100$ &  4.4 & 2.19 & 14.55  & 14.82 & $+0.70 \pm 0.01$ & $-0.30 \pm 0.01$ & $+17.7 \pm 0.1$ & $-0.36 \pm 0.01$ &  1  \\
Berkeley 20  &  1 & 203.4853 & $-17.3763$ &  1.8 & 4.79 & 15.99  & 16.26 & $+0.91 \pm 0.01$ & $-0.27 \pm 0.01$ & $+76.6 \pm 0.2$ & $-0.43 \pm 0.01$ &  1  \\
Berkeley 21  &  1 & 186.8174 & $ -2.4901$ &  3.7 & 2.14 & 14.39  & 14.66 & $+0.46 \pm 0.03$ & $-1.02 \pm 0.02$ & $ +0.5 \pm 1.1$ & $-0.23 \pm 0.05$ &  8  \\
Berkeley 22  &  1 & 199.8736 & $ -8.0708$ &  2.6 & 2.45 & 13.95  & 14.23 & $+0.62 \pm 0.03$ & $-0.40 \pm 0.02$ & $+94.9 \pm 0.8$ & $-0.33 \pm 0.04$ &  6  \\
Berkeley 29  &  1 & 197.9472 & $  7.9816$ &  1.7 & 3.09 & 20.24  & 20.51 & $+0.11 \pm 0.02$ & $-1.05 \pm 0.02$ & $+25.3 \pm 0.1$ & $-0.53 \pm 0.02$ &  2  \\
Berkeley 31  &  1 & 206.2398 & $  5.1334$ &  3.7 & 2.82 & 14.75  & 15.02 & $+0.24 \pm 0.03$ & $-0.89 \pm 0.02$ & $+58.8 \pm 0.9$ & $-0.43 \pm 0.02$ &  2  \\
Berkeley 33  &  1 & 225.4474 & $ -4.5998$ &  3.8 & 0.23 & 12.79  & 13.05 & $-0.69 \pm 0.01$ & $+1.59 \pm 0.01$ & $+77.8 \pm 0.1$ & $-0.24 \pm 0.01$ &  1 \\ 
\multicolumn{13}{c}{......}
\enddata
\tablenotetext{a}{Cluster Radius and age from \citet{cg20}}\vskip-0.07in
\tablenotetext{b}{Calculated with distances from \citet{cg20}, recomputed to a solar radius of $R_0 = 8.274$ kpc.}\vskip-0.07in
\tablenotetext{c}{$\mu_{\alpha}$ and $\mu_{\delta}$ and their $1 \sigma $ uncertainties are those of the 2D Gaussian fit, as in OCCAMII.}\vskip-0.07in
\tablenotetext{}{(This table is available in its entirety in machine-readable form.)}
\end{deluxetable*}


\begin{deluxetable*}{lrrrrrrrr}
\tabletypesize{\tiny}
\tablecaption{OCCAM DR16 Sample - Detailed Chemistry \label{tab:full_sample2}}
	\tablehead{
    \colhead{Cluster} & 
    \colhead{[Fe/H]} &
    \colhead{[O/Fe]} & 
    \colhead{[Na/Fe]} & 
    \colhead{[Mg/Fe]} & 
    \colhead{[Al/Fe]} & 
    \colhead{[Si/Fe]} & 
    \colhead{[S/Fe]} &
    \colhead{[K/Fe]} \\[-5ex]
     \colhead{name} &
     \colhead{(dex)} &
     \colhead{(dex)} &
     \colhead{(dex)} &
     \colhead{(dex)} &
     \colhead{(dex)} &
     \colhead{(dex)} &
     \colhead{(dex)} &
     \colhead{(dex)} \\[-2ex]
    \colhead{} &
    \colhead{[Ca/Fe]} &
    \colhead{[Ti/Fe]} &
    \colhead{[V/Fe]} &
    \colhead{[Cr/Fe]} & 
    \colhead{[Mn/Fe]} & 
    \colhead{[Co/Fe]} & 
    \colhead{[Ni/Fe]} & 
    \colhead{[Ce/Fe]}\\[-5ex] 
     \colhead{} &
     \colhead{(dex)} &
     \colhead{(dex)} &
     \colhead{(dex)} &
     \colhead{(dex)} &
     \colhead{(dex)} &
     \colhead{(dex)} &
     \colhead{(dex)} &
     \colhead{(dex)}
    }
\startdata
\multicolumn{9}{c}{High Quality Clusters} \\ \hline
Berkeley 2   & $-0.21 \pm 0.02$ &  $-0.01 \pm 0.05$ &  $+0.06 \pm 0.66$ &  $-0.02 \pm 0.04$ & $+0.02 \pm 0.03$ & $-0.01 \pm 0.04$ & $+0.08 \pm 0.11$ &  $+0.03 \pm 0.08$ \\ & $+0.03 \pm 0.04$ &  $-0.01 \pm 0.06$ &  $+0.06 \pm 0.27$ &  $+0.04 \pm 0.07$ &  $-0.06 \pm 0.04$ &  $-0.52 \pm 0.40$ &  $-0.02 \pm 0.04$ &  $+0.25 \pm 0.15$ \\[1ex]  
Berkeley 17  & $-0.18 \pm 0.03$ &  $+0.08 \pm 0.02$ &  $+0.00 \pm 0.07$ &  $+0.10 \pm 0.02$ & $+0.10 \pm 0.03$ & $+0.04 \pm 0.02$ & $+0.11 \pm 0.05$ &  $+0.10 \pm 0.04$ \\ & $+0.03 \pm 0.03$ &  $+0.04 \pm 0.05$ &  $-0.12 \pm 0.17$ &  $+0.00 \pm 0.05$ &  $-0.02 \pm 0.02$ &  $+0.08 \pm 0.06$ &  $+0.02 \pm 0.01$ &  $-0.09 \pm 0.08$ \\[1ex]  
Berkeley 18  & $-0.37 \pm 0.03$ &  $+0.09 \pm 0.06$ &  $+0.06 \pm 0.18$ &  $+0.11 \pm 0.02$ & $+0.09 \pm 0.05$ & $+0.08 \pm 0.04$ & $+0.13 \pm 0.08$ &  $+0.17 \pm 0.13$ \\ & $+0.05 \pm 0.07$ &  $+0.03 \pm 0.05$ &  $-0.08 \pm 0.33$ &  $-0.05 \pm 0.14$ &  $-0.02 \pm 0.04$ &  $-0.04 \pm 0.37$ &  $+0.00 \pm 0.05$ &  $+0.11 \pm 0.15$ \\[1ex]  
Berkeley 19  & $-0.36 \pm 0.01$ &  $+0.09 \pm 0.02$ &  $+0.02 \pm 0.08$ &  $+0.13 \pm 0.02$ & $+0.11 \pm 0.03$ & $+0.01 \pm 0.02$ & $+0.06 \pm 0.06$ &  $+0.04 \pm 0.05$ \\ & $+0.01 \pm 0.02$ &  $+0.03 \pm 0.03$ &  $-0.40 \pm 0.09$ &  $-0.02 \pm 0.05$ &  $-0.04 \pm 0.02$ &  $-0.00 \pm 0.07$ &  $-0.02 \pm 0.02$ &  $+0.21 \pm 0.07$ \\[1ex]  
Berkeley 20  & $-0.43 \pm 0.01$ &  $+0.10 \pm 0.01$ &  $+0.04 \pm 0.07$ &  $+0.09 \pm 0.01$ & $+0.15 \pm 0.02$ & $+0.09 \pm 0.02$ & $+0.10 \pm 0.05$ &  $+0.11 \pm 0.05$ \\ & $+0.05 \pm 0.02$ &  $+0.03 \pm 0.02$ &  $-0.21 \pm 0.07$ &  $-0.01 \pm 0.04$ &  $-0.03 \pm 0.02$ &  $-0.01 \pm 0.05$ &  $+0.03 \pm 0.02$ &  $+0.06 \pm 0.05$ \\[1ex]  
Berkeley 21  & $-0.23 \pm 0.05$ &  $-0.00 \pm 0.09$ &  $+0.00 \pm 0.14$ &  $+0.08 \pm 0.04$ & $+0.09 \pm 0.04$ & $-0.04 \pm 0.04$ & $+0.02 \pm 0.08$ &  $-0.01 \pm 0.07$ \\ & $+0.02 \pm 0.03$ &  $+0.04 \pm 0.04$ &  $+0.04 \pm 0.18$ &  $-0.06 \pm 0.15$ &  $-0.04 \pm 0.03$ &  $+0.05 \pm 0.18$ &  $-0.02 \pm 0.03$ &  $+0.25 \pm 0.15$ \\[1ex]  
Berkeley 22  & $-0.33 \pm 0.04$ &  $+0.07 \pm 0.03$ &  $+0.15 \pm 0.11$ &  $+0.09 \pm 0.02$ & $+0.16 \pm 0.08$ & $+0.07 \pm 0.02$ & $+0.03 \pm 0.17$ &  $+0.10 \pm 0.11$ \\ & $+0.04 \pm 0.03$ &  $+0.02 \pm 0.04$ &  $-0.03 \pm 0.20$ &  $-0.11 \pm 0.17$ &  $-0.01 \pm 0.04$ &  $+0.07 \pm 0.12$ &  $-0.00 \pm 0.04$ &  $+0.07 \pm 0.12$ \\[1ex]  
Berkeley 29  & $-0.53 \pm 0.02$ &  $+0.12 \pm 0.01$ &  $+0.13 \pm 0.07$ &  $+0.13 \pm 0.02$ & $+0.02 \pm 0.03$ & $+0.03 \pm 0.02$ & $+0.15 \pm 0.06$ &  $+0.08 \pm 0.05$ \\ & $+0.01 \pm 0.02$ &  $+0.08 \pm 0.02$ &  $-0.18 \pm 0.10$ &  $-0.01 \pm 0.04$ &  $+0.02 \pm 0.02$ &  $+0.15 \pm 0.05$ &  $+0.04 \pm 0.02$ &  $+0.11 \pm 0.05$ \\[1ex]  
Berkeley 31  & $-0.43 \pm 0.02$ &  $+0.11 \pm 0.02$ &  $+0.10 \pm 0.08$ &  $+0.11 \pm 0.02$ & $+0.05 \pm 0.08$ & $+0.06 \pm 0.02$ & $+0.08 \pm 0.06$ &  $+0.14 \pm 0.05$ \\ & $+0.04 \pm 0.02$ &  $+0.05 \pm 0.03$ &  $+0.05 \pm 0.15$ &  $-0.00 \pm 0.05$ &  $-0.05 \pm 0.03$ &  $+0.04 \pm 0.11$ &  $-0.02 \pm 0.03$ &  $+0.85 \pm 0.10$ \\[1ex]  
Berkeley 33  & $-0.24 \pm 0.01$ &  $+0.00 \pm 0.01$ &  $+0.13 \pm 0.06$ &  $+0.01 \pm 0.01$ & $+0.02 \pm 0.02$ & $-0.01 \pm 0.01$ & $+0.05 \pm 0.04$ &  $-0.04 \pm 0.04$ \\ & $+0.01 \pm 0.01$ &  $-0.00 \pm 0.02$ &  $-0.18 \pm 0.06$ &  $-0.03 \pm 0.03$ &  $+0.01 \pm 0.01$ &  $-0.01 \pm 0.04$ &  $-0.05 \pm 0.01$ &  $+0.31 \pm 0.05$ \\[1ex]  
\multicolumn{9}{c}{......}\\
\enddata
\tablenotetext{}{(This table is available in its entirety in machine-readable form.}
\end{deluxetable*}

\subsection{Data Access - SDSS Value Added Catalog}\label{sec:vac}

The VAC consists of two FITS tables. {The first, {\tt occam\_cluster-DR17.fits}, is a combination of Table \ref{tab:full_sample} and Table \ref{tab:full_sample2}, providing bulk cluster parameters derived here, PM from {\it Gaia}, as well as RVs and average abundances for 16 reliable chemical species available in APOGEE DR17.} The second table, {\tt occam\_member-DR17.fits}, contains all of the APOGEE stars considered in this analysis (all of the stars that fall within two radii of the cluster center given by \citet{cg20}; $2 \times Radius_{CG}$) and reports the membership probabilities determined by the OCCAM pipelines (for [Fe/H], RV, and PM) as well as the membership probability from \citet{cg20} for convenience.
These four probabilities reported  reflect how far a given stellar parameter is from the fit cluster mean, where a reported probability of $>0.01$ is within $3\sigma$ of the cluster mean. 
In practice these fit distributions are fairly tight (see \citealt{donor_18} for a figure set showing distributions for [Fe/H], PM and RV for 19 clusters), therefore a star falling within $3\sigma$\footnote{{In practice we adopt a threshold of 0.01 for all membership probabilities; see \citet{donor_18, occam_p4} for further discussion.}} of the cluster mean in all 3 parameter spaces is likely to be a cluster member.

We also note that within the VAC, \rgc was calculated with an $R_{\odot}$ of 8 $\textrm{kpc}$, whereas for this work, we recalculated \rgc with a solar radius of $8.275$ $\textrm{kpc}$ to be consistent with \citet{GRAVITY:2021}.

Table \ref{tab:vac} shows all columns available in the occam\_member table. The catalog is available \href{https://www.sdss.org/dr17/data_access/value-added-catalogs/?vac_id=open-cluster-chemical-abundances-and-mapping-catalog}{from sdss.org.}\footnote{The full url is \url{https://www.sdss.org/dr17/data_access/value-added-catalogs/?vac_id=open-cluster-chemical-abundances-and-mapping-catalog}} 

\begin{deluxetable}{ll}[ht!]
\tablecaption{A summary of the individual star data included in the DR17 OCCAM VAC \label{tab:vac}}
\tabletypesize{\tiny}
\tablehead{
    \colhead{Label} & 
    \colhead{Description}
    }
\startdata
CLUSTER & The associated open cluster \\
2MASS ID &   {star} ID from 2MASS survey\\
LOCATION\_ID\tablenotemark{a} & from APOGEE DR16 \\
GLAT &  Galactic latitude\\
GLON &  Galactic longitude\\
FE\_H\tablenotemark{a}  & [Fe/H]\\
FE\_H\_ERR\tablenotemark{a}  & uncertainty in FE\_H \\
VHELIO\_AVG\tablenotemark{a}  &  heliocentric radial velocity\\
VSCATTER\tablenotemark{a}  & scatter in APOGEE RV measurements \\
PMRA\tablenotemark{b} &  proper motion in right ascension\\
PMDEC\tablenotemark{b}  & proper motion in declination \\
PMRA\_ERR\tablenotemark{b}  & uncertainty in PMRA \\
PMDEC\_ERR\tablenotemark{b}  & uncertainty in PMDEC \\
RV\_PROB &  membership probability based on RV (This study)\\
FEH\_PROB &  membership probability based on FE\_H (This study)\\
PM\_PROB\tablenotemark{c} &  membership probability based on PM (This study)\\
CG\_PROB &  membership probability from \citet{cg_18} \\
\enddata
\tablenotetext{a}{Taken directly from APOGEE DR17.}\vskip-0.1in
\tablenotetext{b}{From \gaia EDR3.}\vskip-0.1in
\tablenotetext{c}{{Negative values indicate the star is outside the adopted cluster radius, while `2' indicates the star failed our PM membership analysis, but is a member in \citet{cg20}.}}
\end{deluxetable}\vskip-0.07in

\section{Results} \label{sec:results}
\subsection{The Galactic Metallicity Gradient}\label{sec:FeH}
\begin{figure*}[t!]
 	\begin{center}
 		\vskip-0.3in
     \plotone{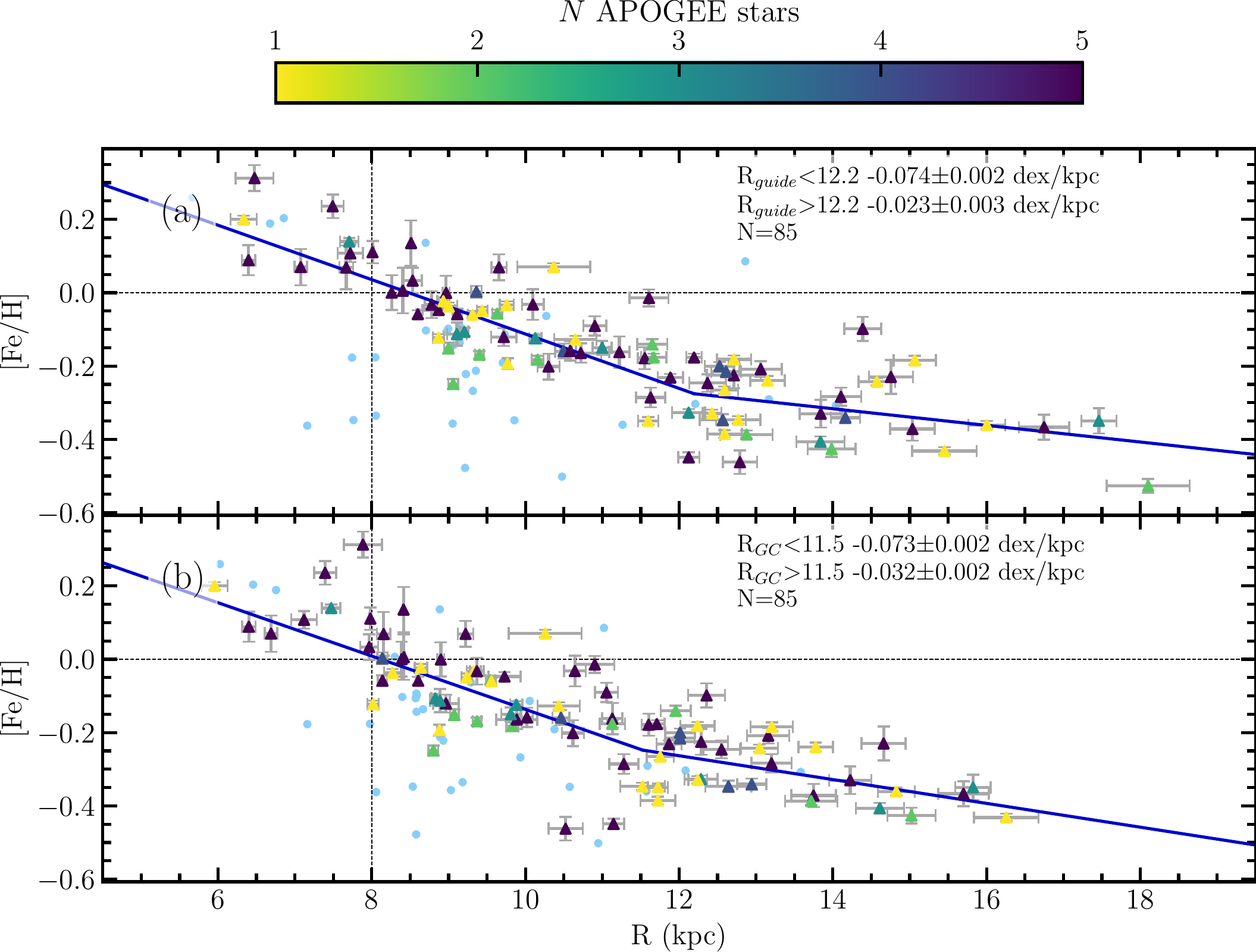} 
 	\end{center}
 	\caption{ \small Metallicity ([Fe/H]) gradients from the full high quality sample mapped as a function of guiding radius (\rguide;  {\em top panel}) and current radius (R$_{GC}$; {\em bottom panel}), along with a bilinear fit as in \citet{occam_p4}. Clusters flagged as potentially unreliable are shown as light blue circles. The color bar indicates the number of OCCAM member stars per cluster, saturating at five.}
 	\label{fig:feh_grad}
 \end{figure*}

With the large, uniform sample of open cluster data from APOGEE DR17, we are well positioned to more reliably characterize and report Galactic abundance gradients for 16 chemical species. 
Figure \ref{fig:feh_grad} shows [Fe/H] versus both \rguide (top panel) and \rgc (bottom panel) for our final sample of 85 open clusters.
In both cases we use a two-function gradient, where the gradient is described with two linear functions and where the intersection point of the two lines is also allowed to be a free parameter. We use the fitting procedures described in OCCAM-IV, {which uses a maximum likelihood method to fit the data, and the {\it emcee} python package \citep{emcee} to estimate the fit errors.  We assume a 5\% error on the distance to each cluster, as \citep{cg20} did not include distance errors, and these are taken into account in the fitting procedure}. 
{We denote the gradient with radius less than the intersection point (hereafter known as the ``knee") as the inner gradient and the gradient with radius greater than the knee as the outer gradient. We find }
an inner gradient of $-0.074 \pm 0.002$ dex/kpc for $R_{guide}$, and a nearly identical inner gradient of $-0.073 \pm 0.002$ for $R_{GC}$. Meanwhile, the outer gradients for the two cases are: d[Fe/H]/\rguide$= -0.023 \pm 0.003$ dex/kpc and d[Fe/H]/$R_{GC} = -0.032 \pm 0.002$ dex/kpc, with the knee located at $12.2 \pm 0.12$ kpc and $11.5 \pm 0.09$ kpc, respectively. 
For completeness, we also fit the open cluster data from Figure \ref{fig:feh_grad} with a single linear function, which is recorded in Table \ref{tab:fehgradients}, along with the two-function fit and the number of clusters used to calculate both fits ($N$).

\begin{deluxetable}{lrrcr}[h!]
\tabletypesize{\scriptsize}
\tablecaption{OCCAM DR17 {[Fe/H]} Gradients \label{tab:fehgradients}}
	\tablehead{
    \colhead{Selection} &
    \colhead{Type} &
    \colhead{Gradient} &
    \colhead{Knee} &
    \colhead{ N} \\[-2ex]
    \colhead{} &
    \colhead{} &
    \colhead{(dex kpc$^{-1}$)} &
    \colhead{(kpc)} &
    \colhead{} \\[-2ex]
    }
	\startdata
\multicolumn{5}{c}{d[Fe/H]/d$R_{GC}$}\\\hline
Inner                        & Knee   & $-0.073 \pm 0.002$ & $11.5 \pm 0.09$  & 85 \\
Outer                        & Knee   & $-0.032 \pm 0.002$ & $11.5 \pm 0.09$  & 85 \\
All                          & Linear & $-0.055 \pm 0.001$ & \nodata          & 85 \\
$\textrm{Age} \le 0.4$       & Linear & $-0.052 \pm 0.003$ & \nodata          & 15 \\ 
$0.4 < \textrm{Age} \le 0.8$ & Linear & $-0.059 \pm 0.003$ & \nodata          & 17 \\
$0.8 < \textrm{Age} \le 2.0$ & Linear & $-0.059 \pm 0.002$ & \nodata          & 29 \\ 
$\textrm{Age} > 2.0$         & Linear & $-0.052 \pm 0.002$ & \nodata          & 22 \\\hline 
 \multicolumn{5}{c}{{d[Fe/H]/d$R_{Guide}$}}\\\hline
Inner                        & Knee   & $-0.074 \pm 0.002$ & $12.2 \pm 0.12$  & 85 \\
Outer                        & Knee   & $-0.023 \pm 0.003$ & $12.2 \pm 0.12$  & 85 \\
All                          & Linear & $-0.056 \pm 0.001$ & \nodata          & 85 \\
$\textrm{Age} \le 0.4$       & Linear & $-0.045 \pm 0.003$ & \nodata          & 15 \\ 
$0.4 < \textrm{Age} \le 0.8$ & Linear & $-0.058 \pm 0.003$ & \nodata          & 17 \\
$0.8 < \textrm{Age} \le 2.0$ & Linear & $-0.065 \pm 0.002$ & \nodata          & 27 \\ 
$\textrm{Age} > 2.0$         & Linear & $-0.049 \pm 0.002$ & \nodata          & 22 
\enddata
\end{deluxetable}

\subsection{Galactic Trends for Other Elements}\label{all_elems}

\subsubsection{$\alpha-$Elements -- O, Mg, Si, S, Ca, Ti}\label{alpha}
\begin{figure}
    \epsscale{1.2}
 	\plotone{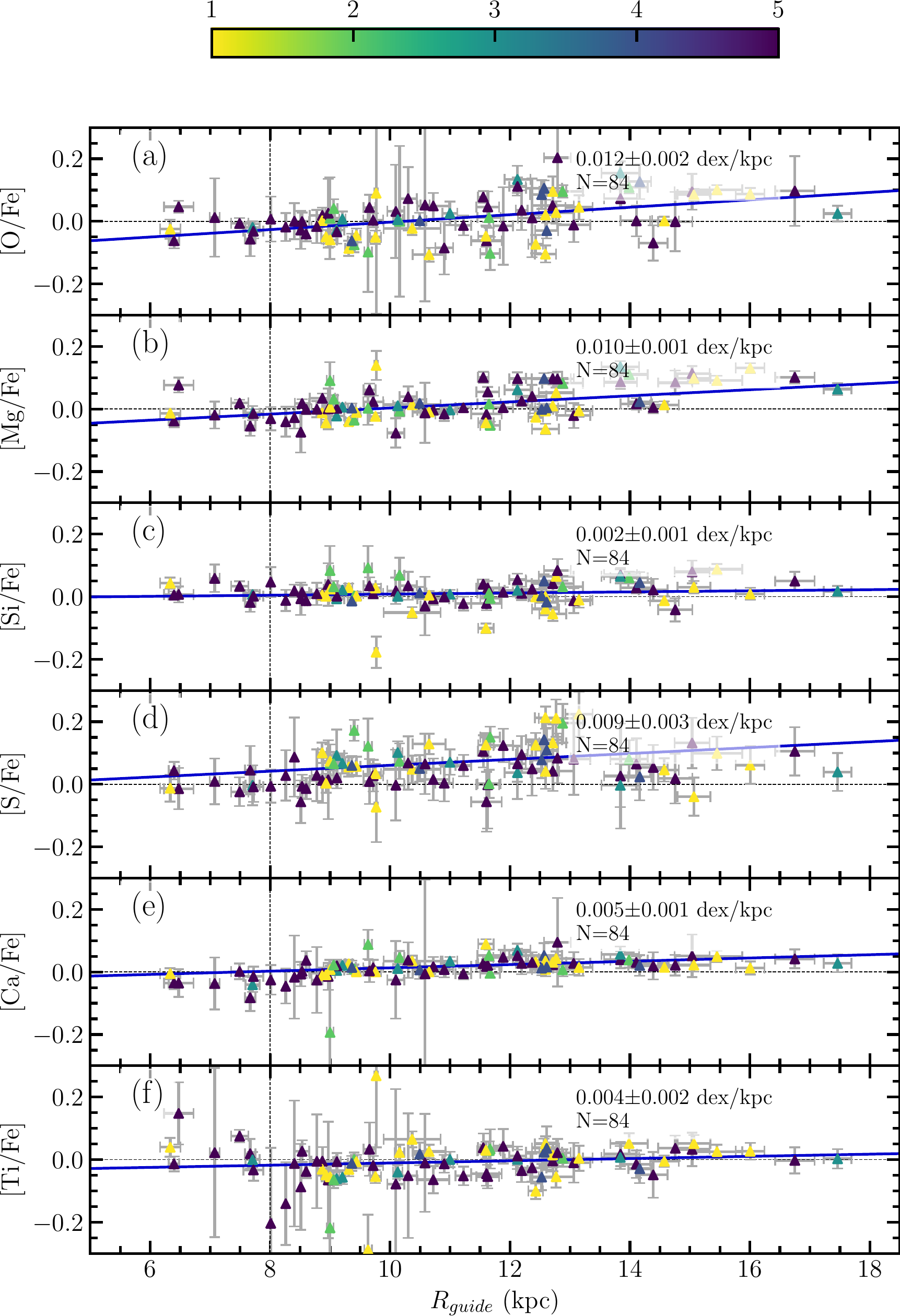}
 	\caption{ \small The [X/Fe] versus \rguide trend for the $\alpha$-elements. As before the color bar indicates number of member stars, saturating at five, and light blue circles represent clusters with high uncertainty in that element.}
 	\label{fig:alpha}
 \end{figure}

The Galactic abundance ratio trends for six $\alpha$-elements (O, Mg, Si, S, Ca, and Ti) over iron versus \rguide are shown in Figure \ref{fig:alpha}, these slopes are also reported in Table \ref{tab:gradients}, along with the slopes calculated with $R_{GC}$. We find positive slopes for all studied [$\alpha$/Fe] abundances but note a significant scatter among the [S/Fe] values and the large uncertainty in the cluster [Ti/Fe] values.
There are no significant differences between the best fit slopes calculated using either \rguide or $R_{GC}$.

\noindent\begin{rotatetable*}
\begin{deluxetable*}{lcrcrcrcrcrcr}
\tabletypesize{\tiny}
\tablecaption{OCCAM DR17 Abundance Gradients \label{tab:gradients}}
	\tablehead{
    \colhead{Age range} &
    \colhead{All Clusters} &
    \colhead{N} &
    \colhead{All Clusters \tablenotemark{a}} &
    \colhead{N} &
    \colhead{$\textrm{Age} \le 0.4$} &
    \colhead{N} &
    \colhead{$0.4 < \textrm{Age} \le 0.8$} &
    \colhead{N} &
    \colhead{$0.8 < \textrm{Age} \le 2.0$} &
    \colhead{N} &
     \colhead{$\textrm{Age} > 2.0$} & 
    \colhead{N} \\[-2.5ex]
    \colhead{$\textrm{R}$ range} &
    \colhead{All $\textrm{R}$} &
    \colhead{} &
    \colhead{$\textrm{R} < 14$ kpc} &
    \colhead{} &
    \colhead{$\textrm{R} < 16$ kpc} &
    \colhead{} &
    \colhead{$\textrm{R} < 16$ kpc} &
    \colhead{} &
    \colhead{$\textrm{R} < 16$ kpc} &
    \colhead{} &
     \colhead{$\textrm{R} < 16$ kpc} & 
    \colhead{} \\[-2.5ex]
    \colhead{Gradient} &
    \colhead{(dex kpc$^{-1}$)} &
    \colhead{} &
    \colhead{(dex kpc$^{-1}$)} &
    \colhead{} &
    \colhead{(dex kpc$^{-1}$)} &
    \colhead{} &
    \colhead{(dex kpc$^{-1}$)} &
    \colhead{} &
    \colhead{(dex kpc$^{-1}$)} &
    \colhead{} &
     \colhead{(dex kpc$^{-1}$)} & 
    \colhead{} \\[-2ex]
    }
	\startdata
\multicolumn{13}{c}{Gradients for $R_{GC}$}\\\hline
{d[O/Fe]/d$R_{GC}$}      & $+0.014 \pm 0.002$ & 84 & $+0.015 \pm 0.002$ & 73 & $-0.000 \pm 0.004$ & 15 & $+0.014 \pm 0.007$ & 17 & $+0.009 \pm 0.004$ & 29 & $+0.010 \pm 0.004$ & 22 \\ 
{d[Mg/Fe]/d$R_{GC}$}     & $+0.011 \pm 0.001$ & 84 & $+0.008 \pm 0.001$ & 73 & $+0.001 \pm 0.003$ & 15 & $+0.002 \pm 0.003$ & 17 & $+0.007 \pm 0.002$ & 29 & $+0.012 \pm 0.003$ & 22 \\ 
{d[Si/Fe]/d$R_{GC}$}     & $+0.002 \pm 0.001$ & 84 & $+0.001 \pm 0.001$ & 73 & $-0.003 \pm 0.003$ & 15 & $-0.013 \pm 0.004$ & 17 & $+0.001 \pm 0.002$ & 29 & $+0.002 \pm 0.003$ & 22 \\ 
{d[S/Fe]/d$R_{GC}$}      & $+0.010 \pm 0.003$ & 84 & $+0.017 \pm 0.004$ & 73 & $+0.012 \pm 0.007$ & 15 & $+0.014 \pm 0.011$ & 17 & $+0.010 \pm 0.006$ & 29 & $+0.013 \pm 0.008$ & 22 \\ 
{d[Ca/Fe]/d$R_{GC}$}     & $+0.005 \pm 0.001$ & 84 & $+0.007 \pm 0.002$ & 73 & $+0.006 \pm 0.003$ & 15 & $+0.011 \pm 0.004$ & 17 & $+0.004 \pm 0.003$ & 29 & $+0.005 \pm 0.004$ & 22 \\ 
{d[Ti/Fe]/d$R_{GC}$}     & $+0.004 \pm 0.002$ & 84 & $+0.003 \pm 0.003$ & 73 & $+0.002 \pm 0.005$ & 15 & $+0.015 \pm 0.008$ & 17 & $-0.003 \pm 0.004$ & 29 & $+0.005 \pm 0.006$ & 22 \\\hline 
{d[V/Fe]/d$R_{GC}$}      & $-0.012 \pm 0.008$ & 64 & $+0.028 \pm 0.011$ & 58 & $+0.037 \pm 0.016$ & 14 & $+0.014 \pm 0.031$ & 17 & $+0.009 \pm 0.017$ & 29 & $-0.024 \pm 0.020$ & 22 \\ 
{d[Cr/Fe]/d$R_{GC}$}     & $-0.003 \pm 0.004$ & 76 & $-0.002 \pm 0.005$ & 65 & $+0.000 \pm 0.008$ & 15 & $+0.006 \pm 0.012$ & 17 & $-0.009 \pm 0.006$ & 29 & $-0.008 \pm 0.008$ & 22 \\ 
{d[Mn/Fe]/d$R_{GC}$}     & $-0.007 \pm 0.002$ & 82 & $-0.019 \pm 0.002$ & 71 & $-0.009 \pm 0.004$ & 14 & $+0.002 \pm 0.004$ & 17 & $-0.008 \pm 0.003$ & 29 & $-0.008 \pm 0.004$ & 22 \\ 
{d[Co/Fe]/d$R_{GC}$}     & $-0.006 \pm 0.005$ & 62 & $-0.015 \pm 0.007$ & 56 & $-0.014 \pm 0.011$ & 12 & $-0.040 \pm 0.027$ & 14 & $-0.014 \pm 0.011$ & 29 & $-0.004 \pm 0.011$ & 22 \\ 
{d[Ni/Fe]/d$R_{GC}$}     & $-0.000 \pm 0.001$ & 84 & $-0.003 \pm 0.002$ & 73 & $-0.007 \pm 0.003$ & 15 & $+0.004 \pm 0.004$ & 17 & $-0.003 \pm 0.002$ & 29 & $-0.002 \pm 0.003$ & 22 \\\hline  
{d[Na/Fe]/d$R_{GC}$}     & $-0.021 \pm 0.006$ & 66 & $-0.031 \pm 0.008$ & 56 & $-0.040 \pm 0.014$ & 15 & $-0.021 \pm 0.022$ & 17 & $-0.025 \pm 0.011$ & 29 & $-0.014 \pm 0.013$ & 22 \\ 
{d[Al/Fe]/d$R_{GC}$}     & $+0.009 \pm 0.002$ & 82 & $+0.005 \pm 0.003$ & 71 & $-0.001 \pm 0.005$ & 15 & $-0.005 \pm 0.005$ & 17 & $+0.008 \pm 0.004$ & 29 & $+0.008 \pm 0.006$ & 22 \\ 
{d[K/Fe]/d$R_{GC}$}      & $+0.017 \pm 0.003$ & 80 & $+0.017 \pm 0.004$ & 69 & $+0.003 \pm 0.008$ & 14 & $+0.027 \pm 0.011$ & 17 & $+0.015 \pm 0.006$ & 29 & $+0.003 \pm 0.009$ & 22 \\\hline  
{d[Ce/Fe]/d$R_{GC}$}     & $+0.022 \pm 0.006$ & 69 & $+0.044 \pm 0.009$ & 60 & $+0.034 \pm 0.016$ & 12 & $+0.045 \pm 0.027$ & 14 & $+0.035 \pm 0.012$ & 29 & $+0.056 \pm 0.014$ & 21 \\\hline  
\multicolumn{13}{c}{Gradients for $R_{guide}$}\\\hline
{d[O/Fe]/d$R_{guide}$}   & $+0.012 \pm 0.002$ & 84 & $+0.013 \pm 0.002$ & 73 & $+0.002 \pm 0.003$ & 15 & $+0.016 \pm 0.008$ & 17 & $+0.011 \pm 0.005$ & 27 & $+0.008 \pm 0.003$ & 22 \\ 
{d[Mg/Fe]/d$R_{guide}$}  & $+0.010 \pm 0.001$ & 84 & $+0.008 \pm 0.001$ & 73 & $+0.002 \pm 0.003$ & 15 & $+0.005 \pm 0.003$ & 17 & $+0.003 \pm 0.002$ & 27 & $+0.010 \pm 0.002$ & 22 \\ 
{d[Si/Fe]/d$R_{guide}$}  & $+0.002 \pm 0.001$ & 84 & $+0.000 \pm 0.002$ & 73 & $-0.002 \pm 0.002$ & 15 & $-0.013 \pm 0.004$ & 17 & $-0.000 \pm 0.003$ & 27 & $+0.007 \pm 0.003$ & 22 \\ 
{d[S/Fe]/d$R_{guide}$}   & $+0.009 \pm 0.003$ & 84 & $+0.018 \pm 0.004$ & 73 & $+0.009 \pm 0.006$ & 15 & $+0.016 \pm 0.012$ & 17 & $+0.013 \pm 0.007$ & 27 & $+0.013 \pm 0.007$ & 22 \\ 
{d[Ca/Fe]/d$R_{guide}$}  & $+0.005 \pm 0.001$ & 84 & $+0.008 \pm 0.002$ & 73 & $+0.005 \pm 0.003$ & 15 & $+0.009 \pm 0.004$ & 17 & $+0.005 \pm 0.003$ & 27 & $+0.007 \pm 0.003$ & 22 \\ 
{d[Ti/Fe]/d$R_{guide}$}  & $+0.004 \pm 0.002$ & 84 & $+0.002 \pm 0.003$ & 73 & $+0.003 \pm 0.004$ & 15 & $+0.014 \pm 0.008$ & 17 & $-0.006 \pm 0.005$ & 27 & $+0.004 \pm 0.006$ & 22 \\\hline  
{d[V/Fe]/d$R_{guide}$}   & $-0.011 \pm 0.008$ & 64 & $+0.038 \pm 0.011$ & 58 & $+0.024 \pm 0.014$ & 14 & $+0.037 \pm 0.031$ & 17 & $+0.013 \pm 0.018$ & 27 & $-0.016 \pm 0.018$ & 22 \\ 
{d[Cr/Fe]/d$R_{guide}$}  & $-0.003 \pm 0.003$ & 76 & $-0.001 \pm 0.005$ & 65 & $-0.000 \pm 0.007$ & 15 & $+0.008 \pm 0.013$ & 17 & $-0.006 \pm 0.008$ & 27 & $-0.007 \pm 0.008$ & 22 \\ 
{d[Mn/Fe]/d$R_{guide}$}  & $-0.007 \pm 0.002$ & 82 & $-0.011 \pm 0.002$ & 71 & $-0.007 \pm 0.003$ & 14 & $+0.002 \pm 0.005$ & 17 & $-0.012 \pm 0.003$ & 27 & $-0.008 \pm 0.004$ & 22 \\ 
{d[Co/Fe]/d$R_{guide}$}  & $-0.007 \pm 0.005$ & 62 & $-0.023 \pm 0.007$ & 56 & $-0.010 \pm 0.009$ & 12 & $-0.069 \pm 0.033$ & 14 & $-0.038 \pm 0.014$ & 27 & $-0.008 \pm 0.009$ & 22 \\ 
{d[Ni/Fe]/d$R_{guide}$}  & $-0.001 \pm 0.001$ & 84 & $-0.004 \pm 0.002$ & 73 & $-0.006 \pm 0.002$ & 15 & $+0.003 \pm 0.004$ & 17 & $-0.006 \pm 0.003$ & 27 & $+0.001 \pm 0.003$ & 22 \\\hline  
{d[Na/Fe]/d$R_{guide}$}  & $-0.020 \pm 0.006$ & 66 & $-0.035 \pm 0.008$ & 56 & $-0.030 \pm 0.012$ & 15 & $-0.022 \pm 0.022$ & 17 & $-0.032 \pm 0.013$ & 27 & $-0.013 \pm 0.010$ & 22 \\ 
{d[Al/Fe]/d$R_{guide}$}  & $+0.009 \pm 0.002$ & 82 & $+0.005 \pm 0.003$ & 71 & $+0.001 \pm 0.004$ & 15 & $-0.001 \pm 0.006$ & 17 & $+0.008 \pm 0.004$ & 27 & $+0.008 \pm 0.005$ & 22 \\ 
{d[K/Fe]/d$R_{guide}$}   & $+0.016 \pm 0.003$ & 80 & $+0.017 \pm 0.004$ & 69 & $+0.004 \pm 0.007$ & 14 & $+0.029 \pm 0.011$ & 17 & $+0.014 \pm 0.008$ & 27 & $+0.007 \pm 0.008$ & 22 \\\hline  
{d[Ce/Fe]/d$R_{guide}$}  & $+0.024 \pm 0.006$ & 69 & $+0.051 \pm 0.010$ & 60 & $+0.029 \pm 0.013$ & 12 & $+0.047 \pm 0.027$ & 14 & $+0.050 \pm 0.014$ & 27 & $+0.028 \pm 0.012$ & 21 \\\hline  
\enddata
\tablenotetext{a}{{While not explicitly discussed in text we report the gradient cut at 14kpc in order to easily compare to previous work.}}\vskip-0.1in
\end{deluxetable*}
\end{rotatetable*}

\subsubsection{Iron-Peak Elements -- V, Cr, Mn, Co, Ni}\label{ironpeak}

\begin{figure}[t!]
    \begin{center}
    \epsscale{1.1}
 	\plotone{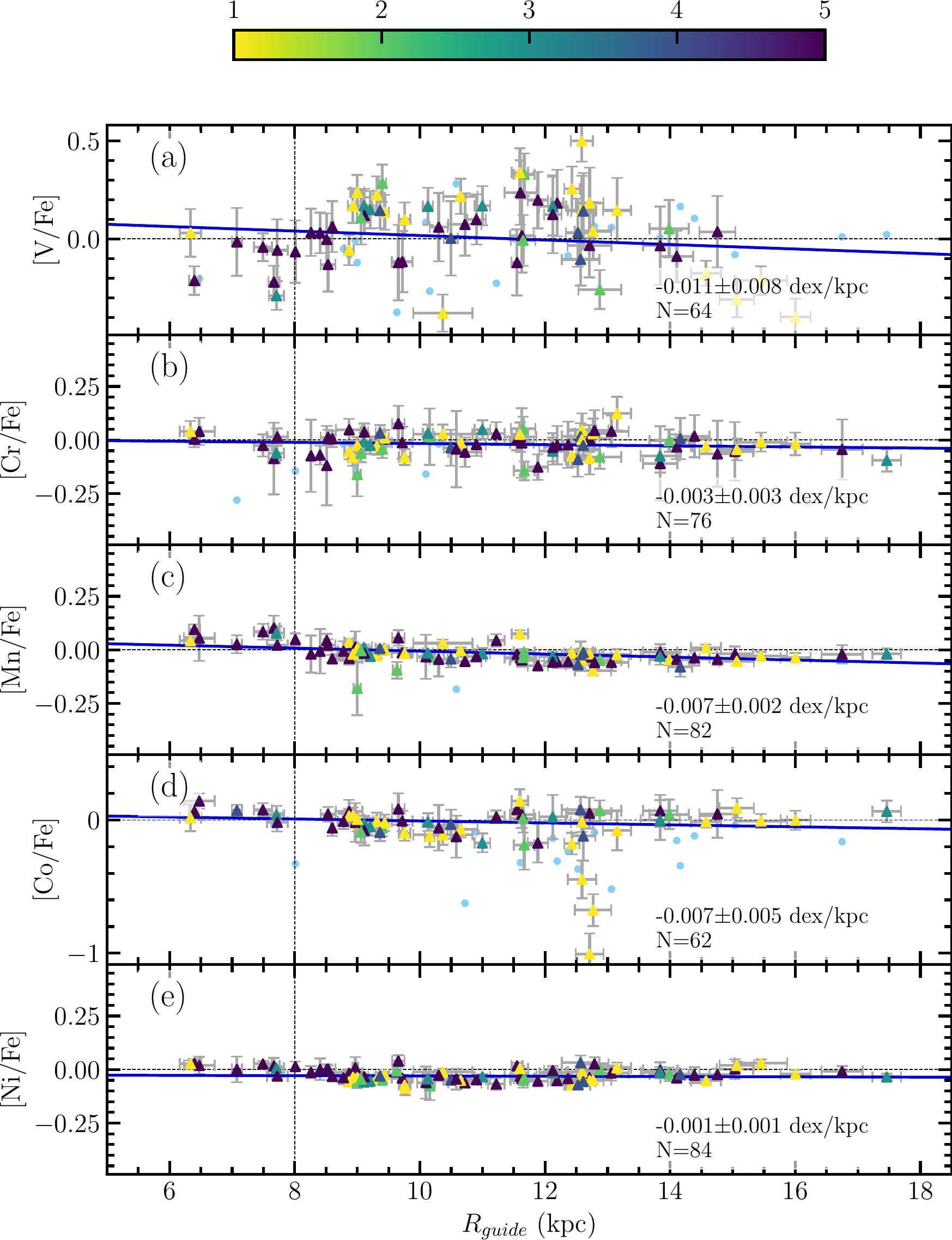}
 	\vskip0.1in
 	\caption{ \small Same as Figure \ref{fig:alpha}, but for the iron-peak elements.}
 	\label{fig:ironpeak}
 	\end{center}
\end{figure}

In Figure \ref{fig:ironpeak}, we investigate the Galactic trends versus \rguide of the iron-peak element ratios included in DR17 (V, Cr, Mn, Co, and Ni).\footnote{As discussed in \citet[][]{dr17} and Holtzman et al.\ ({\it in prep}), the APOGEE DR17 pipeline analysis did not yield sufficiently reliable abundance measurements for the element copper.} 
The gradients for iron-peak elements over iron all show negative, shallow trends (Table \ref{tab:gradients})
with vanadium having the steepest gradient value of all at $-0.012 \pm 0.008$ dex/kpc, although this is still relatively flat. The cluster values for [V/Fe] also have the largest scatter of the iron-peak elements, 
however, [Co/Fe] also has three significant outliers: the clusters FSR 0716 ([Co/Fe] $= -0.45$ dex), FSR 1113 ([Co/Fe] $= -0.68$ dex) and Haffner 4 ([Co/Fe] $= -1.01$ dex), which all only have one stellar member in our sample.

\subsubsection{Odd-Z Elements -- Na, Al, K}
\begin{figure}
    \epsscale{1.1}
 	\plotone{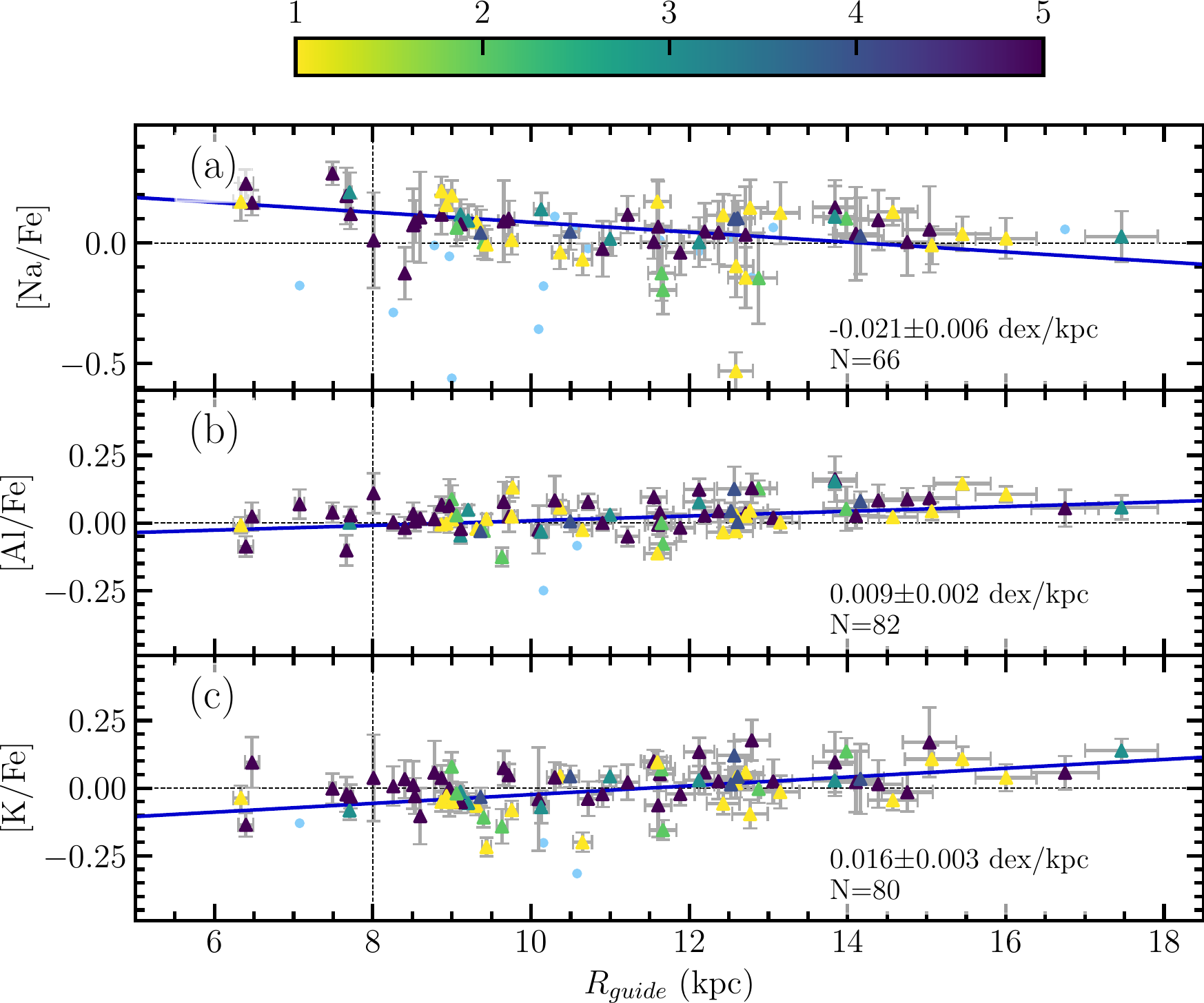}
 	\caption{ \small Same as Figure \ref{fig:alpha} but for the ``odd-z" elements. 
    }
 	\label{fig:random}
\end{figure}

 The abundance gradients with respect to \rguide for the three ``odd-z'' elements: Na, Al, and K are plotted in Figure \ref{fig:random}, and recorded in Table \ref{tab:gradients} for both \rguide and $R_{GC}$. We report similar positive trends for Al and K, but a steep negative gradient for Na. The single one-star outlier for [Na/Fe] corresponds to the open cluster NGC 136.

\subsubsection{The Neutron Capture Element Ce}\label{Ce}

\begin{figure}[t!]
    \begin{center}
    \epsscale{1.2}
 	\plotone{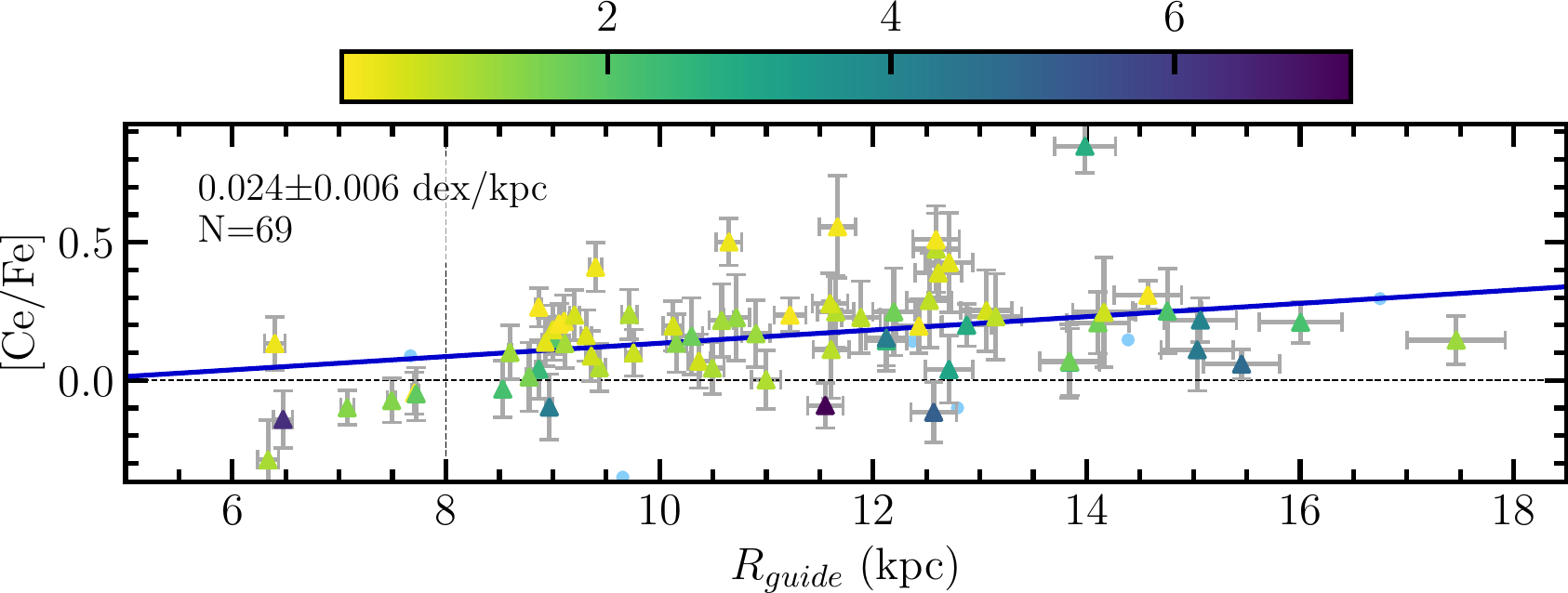}
 	\vskip0.1in
 	\caption{ \small The Galactic abundance trend for cerium. The points are colored by their age in Gyrs. }
 	\label{fig:neutron}
 	\end{center}
\end{figure}

{With the availability of reliable abundances for the s-process element Ce, obtained automatically by the ASPCAP pipeline in DR17, we are now able to investigate the abundance gradient of Ce with 69 open clusters in our sample.} 
In Figure \ref{fig:neutron}, we fit [Ce/Fe] abundance versus \rguide and find a positive gradient of $0.024 \pm 0.006$ dex/kpc. In Table \ref{tab:gradients}, we also report the slope with respect to \rgc and find a value of $0.022 \pm 0.006$ dex/kpc.

\subsection{The Evolution of Galactic Abundance Gradients}
\subsubsection{Iron}\label{sec:fe_evo}
One of the important questions in chemical evolution models is how the Galactic metallicity gradients have evolved over time.  Fortunately, the size of our sample lends itself to investigating this question.
The open cluster sample studied here can be split into four age bins, divided at 400 Myrs, 800 Myrs, and 2 Gyrs, identical to the bins chosen in OCCAM-IV, although in this study we use the open cluster ages derived in \citet{cg20}.

\begin{figure}[ht!]
    \epsscale{1.2}
 	\plotone{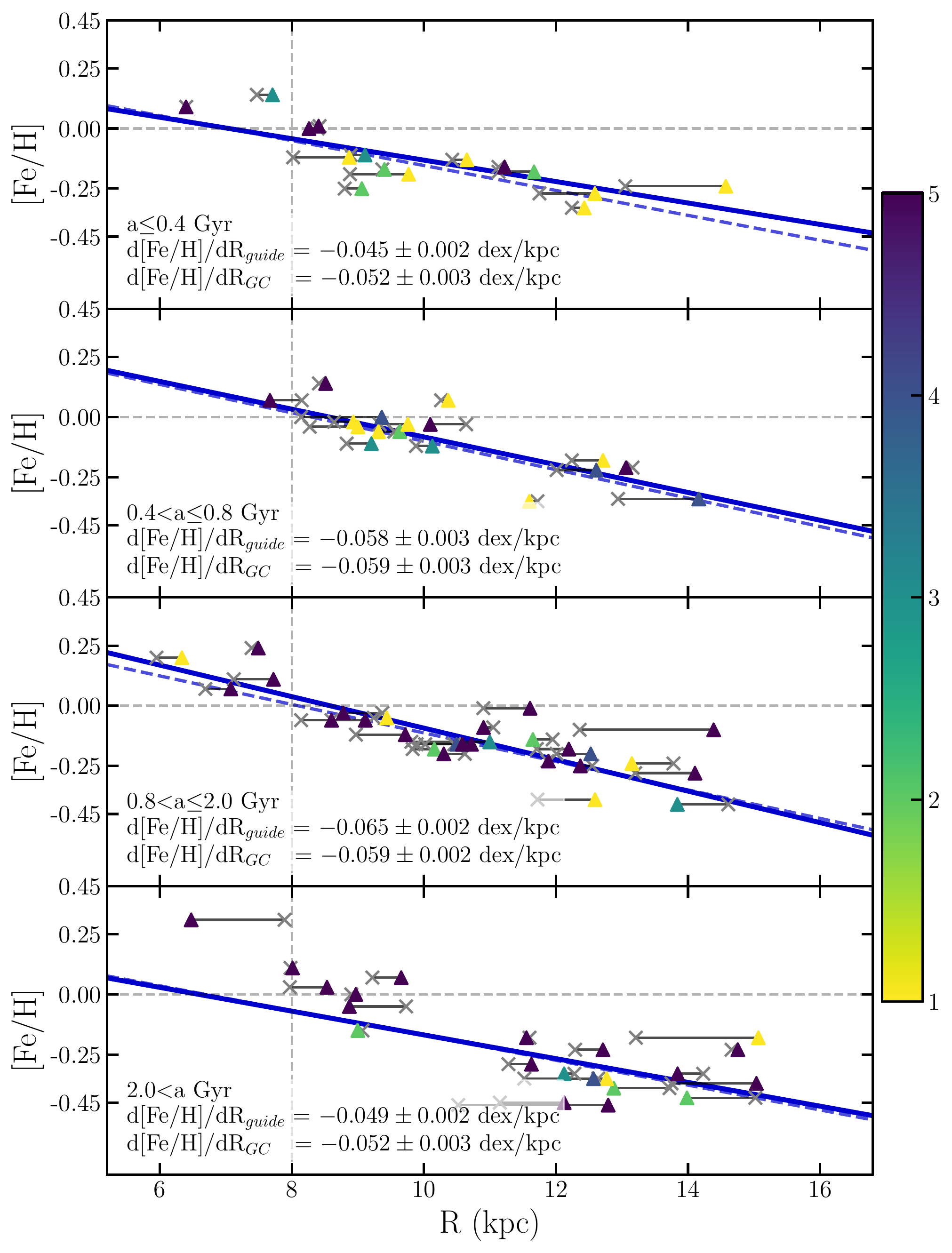}
 	\caption{ \small The Galactic [Fe/H] versus radius trend in four age bins. Gray `X's represent the \rgc of the cluster, while the colored triangles show $R_{guide}$, both of these values are connected with a thin grey line for each cluster. The solid line shows the [Fe/H] versus \rguide trend, and the dashed line is the trend for $R_{GC}$.}
 	\label{fig:comprad}
\end{figure}

In Figure \ref{fig:comprad}, we plot \rguide and \rgc versus metallicity for each age bin, showing only clusters with {both \rgc and \rguide $< 16$ kpc. The gradients shown in the figure, however, are calculated with all clusters located within that region (thus the number of clusters changes between the \rgc and \rguide fits).}
This gives a sample of 73 clusters for the \rguide plots and a sample of 76 clusters for the $R_{GC}$.
Two sets of symbols are used in Figure \ref{fig:comprad}: colored triangles denote guiding center radii while galactocentric radii are marked with gray Xs; horizontal bars connect the two radii values for the same cluster.
The slope of the gradient calculated with respect to \rguide is shown as a solid line, and the slope calculated with \rgc is represented as a dashed line. The slopes calculated for each age bin and the number of clusters used for each fit are recorded in Table \ref{tab:fehgradients}. 

{
As can be seen in Figure \ref{fig:comprad}, the gradients calculated with \rgc appear to remain relatively constant between the four age bins, with the first and the fourth bins showing relatively shallow slopes and the two middle bins having identical slopes. The gradients calculated with \rguide seem to show a more constant transition from young to old clusters up until the final age bin, wherein the slope becomes significantly shallower. 
Additionally, we can see in the last two panels of Figure \ref{fig:comprad} (i.e,. the two older age bins), that on average the difference between a cluster's \rgc and \rguide is larger than in the first two age bins (i.e., the younger two age bins).  This suggests that as the clusters have had more time to be affected by interactions in the Galaxy, e.g asymmetric drift, their orbits have become more elliptical. Plus, a potential survivor bias in the older cluster samples and/or possible radial migration of clusters could have affected the gradients. }

\subsubsection{[X/Fe]}\label{elem_evo}
\begin{figure*}[t!]
    \epsscale{1.0}
 	\plotone{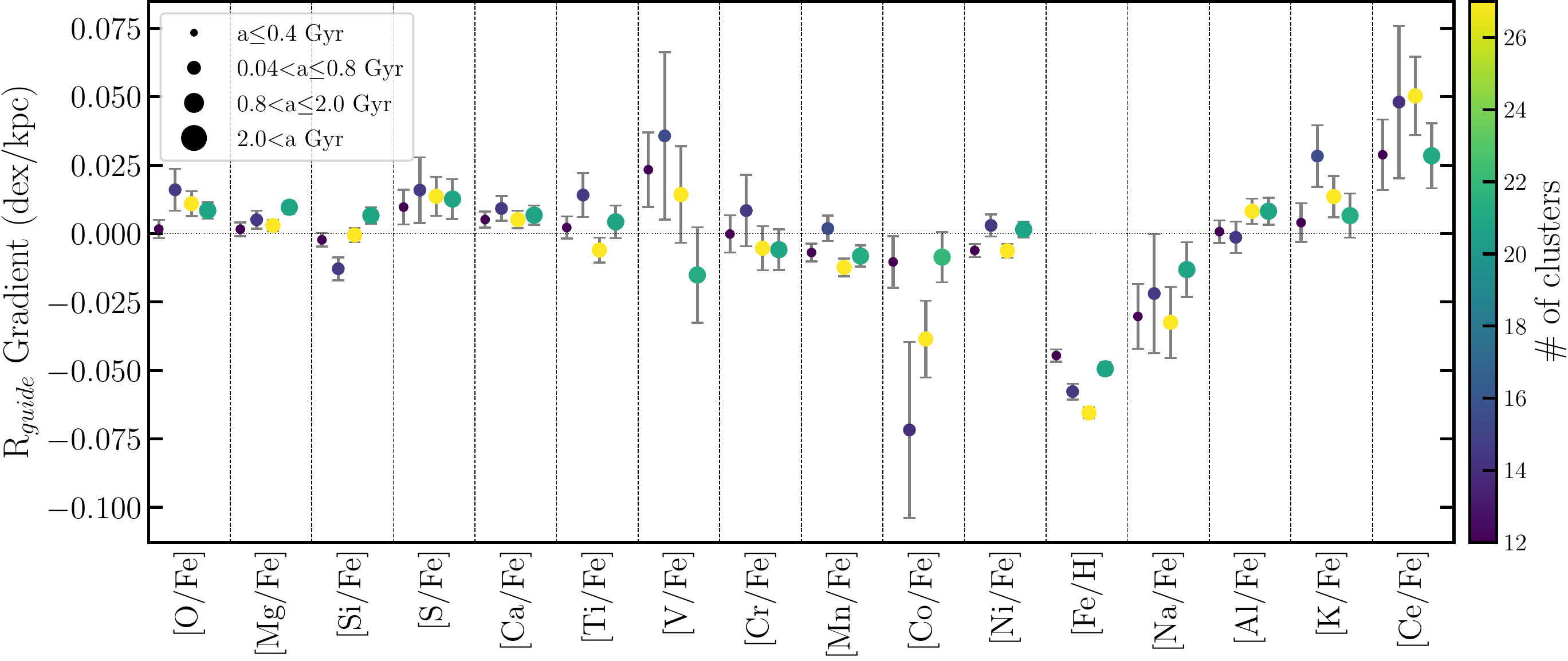}
 	\caption{ \small The slopes of each elemental gradient in four age bins (as in Figure \ref{fig:comprad}), for $R_{guide}$. Point size increases with age, and the color indicates the number of clusters included in the gradient measurement. }
 	\label{fig:slopesummary}
 \end{figure*}

To understand the evolution in the radial gradients of elements other than iron, we split the cluster sample into the same four age bins as in \S \ref{sec:fe_evo} and fit each gradient as in Figure \ref{fig:comprad}. Fit parameters for all elements measured both with \rguide and with \rgc are reported in Table \ref{tab:gradients}.
In Figure \ref{fig:slopesummary}, we also show the slopes ($d[{\rm X/Fe}]/dR_{guide}$) for all four age bins and each of the 16 elements (where, for iron, we show the abundance ratio [Fe/H]); this figure is comparable to Figure 14 in OCCAM-IV. We note that, as explained in \S \ref{sec:fe_evo}, all clusters used in the fit have a radius (\rguide or $R_{GC}$) less than 16 kpc. 

We find no convincing trends through the four age bins in $\alpha$ elements. While there could be a slight trend in [Mg/Fe], with oldest clusters perhaps showing a steeper slope than younger clusters, the changes between samples are roughly as significant as the uncertainties.

{Both [Cr/Fe] and [Ni/Fe] hover around a flat gradient throughout all four age bins, and the gradients of [V/Fe] and [Cr/Fe] both have large uncertainties in their measurements, which makes it difficult to determine any evolutionary trends. }
Additionally, we do not find a significant trend for [Mn/Fe], which breaks with previous APOGEE-based DR16 results presented in OCCAM-IV.

In the odd-Z elements, the gradient for [Na/Fe] seems to have an increasingly negative trend in \rguide as clusters get younger, though within the sizeable uncertainty the trend may be less significant.
Finally for cerium, the uncertainties in the DR17 measurements are still too large to measure a significant trend over time.

\section{Discussion} \label{sec:discussion}
    
\begin{figure}[h!]
    \epsscale{1.2}
    \plotone{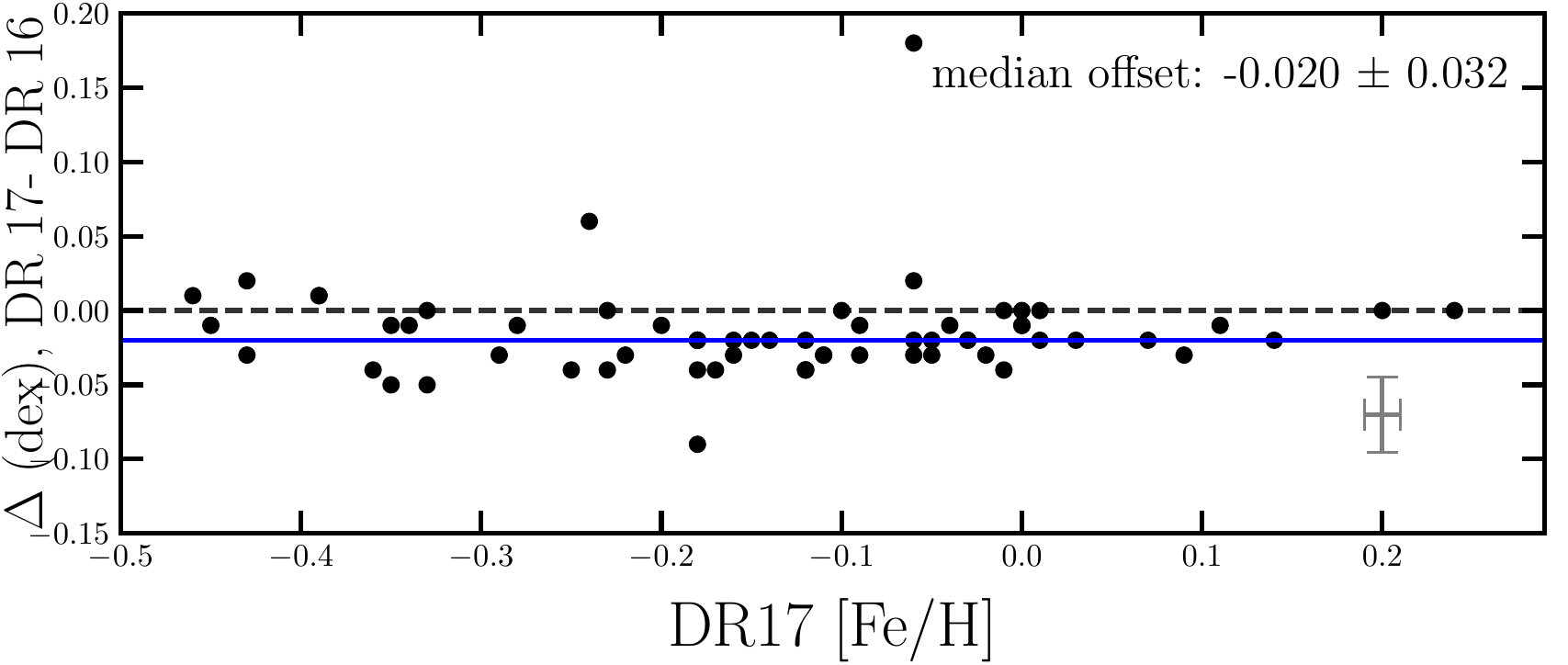}
 	\caption{ \small Comparing the DR17 and DR16 [Fe/H] abundances. The measured median offset is shown as a solid blue line. A characteristic error bar is shown for reference.}
 	\label{fig:compFe}
\end{figure}

\subsection{Comparison to OCCAM-IV sample\label{sec:dr16comp}}

Between this sample and OCCAM-IV, 111 clusters can be found in both samples, 42 new clusters were added to this sample, and 17 clusters were not recovered, including two ``high quality'' clusters: Berkeley 44 and NGC 2355.  
With updated \gaia EDR3 data, the 2D Gaussian fit to the kernel convolution in proper motion space was narrower by enough that the APOGEE star now fell further than $3\sigma$ from the distribution. 
{For Berkeley 44, the star that was included in OCCAM-IV is now slightly outside of the 2D Gaussian fit to the \gaia EDR3 proper motions. It is also not reported as a member in CG18.
For NGC 2355, the star that was included in OCCAM-IV is considered a member in CG18 with a membership probability of 70\%, but using updated EDR3 proper motions, the 2D Gaussian fit was more narrow and thus this star was rejected by our pipeline. }

Additionally, there were three ``high-quality'' clusters in OCCAM-IV which were demoted to being flagged as ``potentially unreliable" in this sample (SAI 16, BH 211, and Basel 11b). BH 211 failed the visual quality check, and both Basel 11b and SAI 16 had only two potential members with conflicting [Fe/H] values. However, there are seven clusters (Berkeley 91, FSR 0496, King 8, NGC 136, NGC 2202, Saurer 1, and Teutsch 10) which were previously marked ``0'' or ``potentially unreliable'' that are now included in the ``high quality'' sample due to the addition of new data.

For designated ``high quality'' clusters in common between this sample and \citet{occam_p4}, a total of 66 clusters, Figure \ref{fig:compFe} shows the change in [Fe/H] between APOGEE DR16 and DR17. The median change, measured to be -0.020, is well within the measured scatter of 0.033, although this scatter seems to be due mostly to the lowest metallicity clusters ([Fe/H] $\lesssim -0.4$). A visual inspection of the plot suggests that closer to Solar [Fe/H] there may be a real, albeit slight offset from DR16. However, this small offset is easily explainable by the significant changes to the APOGEE pipeline. The single outlier in Figure \ref{fig:compFe} with a $\Delta$[Fe/H]$= 0.18$ is NGC 752.

\begin{figure*}
    \epsscale{1.2}
    \plotone{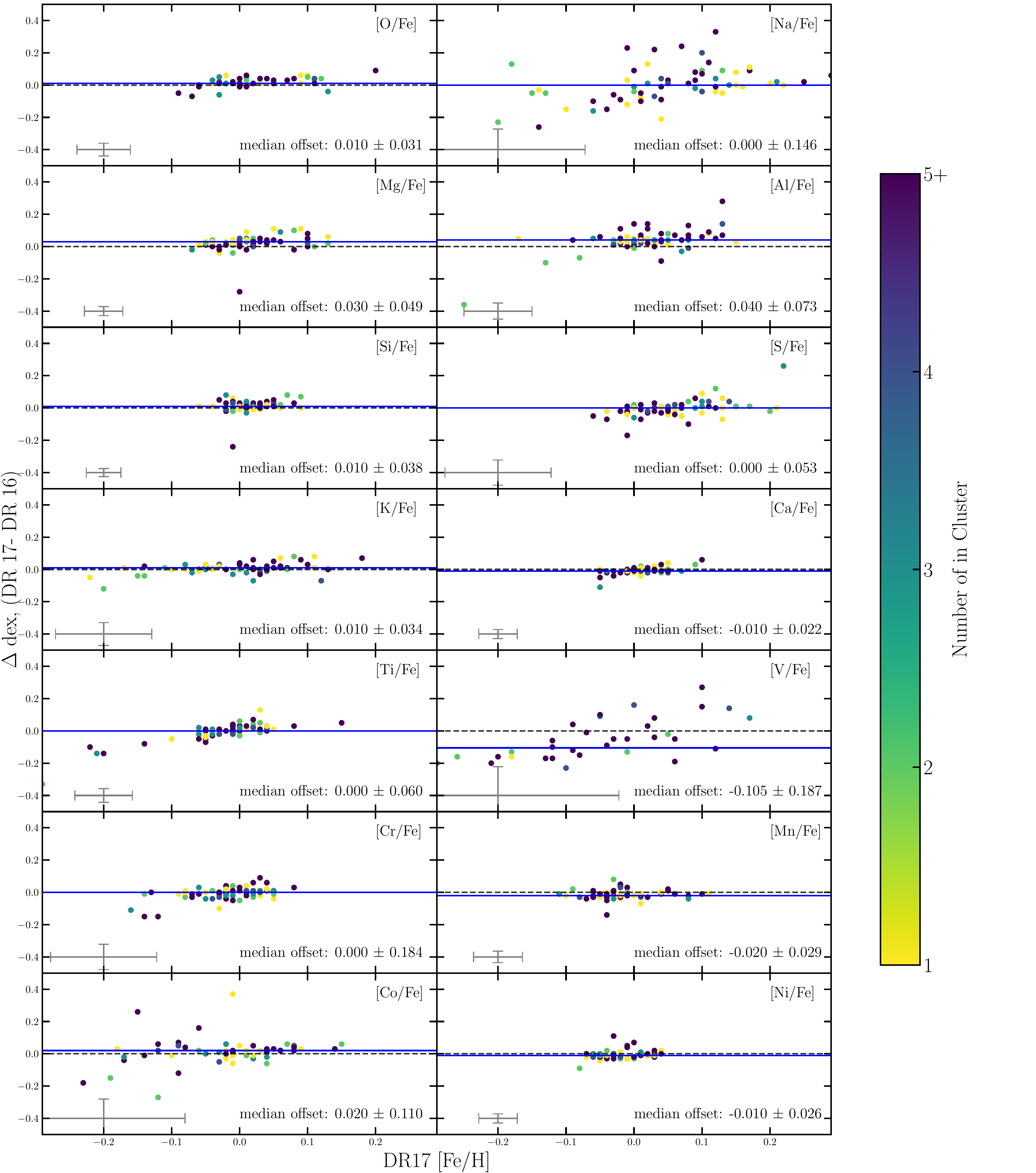}
	\caption{ \small A comparison of DR17 versus DR16 for each chemical element in our study. Characteristic error bars, calculated identically to those in Figure \ref{fig:compFe}, are indicated in each panel. Median offsets between the DR17 and DR16 values for clusters in common are shown by a blue line, while datapoints are colored by the number of stars in the DR17 cluster.}

	\label{fig:dr16v17}
\end{figure*}

Figure \ref{fig:dr16v17} shows the change in \occam measured cluster abundances for 14 elements from APOGEE DR16 to DR17, plotted as a function of their reported DR17 abundance. These differences are due to pipeline and membership changes. Copper and phosphorus are not included because of unsuccessful measurements in DR17 \citep[][Holtzman et. al, \textit{in prep}]{dr17}. Cerium is also not included in Figure \ref{fig:dr16v17} because the values reported in DR16 were not considered particularly reliable \citep{jonsson_2020}. The measured median offset is within the measured scatter for all 14 abundances investigated. It is worth commenting on the particularly large scatter, and potential trend, for vanadium and sodium. Vanadium is considered less reliable in both DR16 and DR17; sodium is considered reliable in DR17 but less so in DR16 \citep{jonsson_2020}.

\subsection{Comparison to other surveys} \label{sec:other_surveys}

\citet{spina21} use data from GALAH+, APOGEE DR16, and \gaia to compile a list of 226 open clusters, 134 of which have high-quality spectroscopic data for up to 21 elements. Of these clusters 85 are in common with our sample. We compare our sample to the GALAH sample, much like Figure \ref{fig:compFe}, and measure a median offset, ($\Delta$ dex, DR17-GALAH) of $-0.018 \pm 0.046$, 
with two major outliers: King 2 at $+0.18$ and Berkeley 18 at $+0.25$, which both only have one member in the GALAH catalog.

In a recent APOGEE study, \citet{apogee_Ce} investigated the abundance gradient for the s-process element, Ce, with a detailed abundance analysis of several Ce II lines from \citet{Cunha_17}. They use 218 stellar members of 42 open clusters from the OCCAM-IV sample. In a manner identical to the comparisons above for the OCCAM-IV and GALAH surveys, we compare the [Ce/Fe] abundances for all clusters in common between this sample and the one reported in \citet{apogee_Ce}. We find not only a systematic shift, but also a sub-solar offset for the cerium abundances between the two samples, both between the open clusters and individual stellar abundances. This shift may be due to BACCHUS, as used by \citet{apogee_Ce}, not properly excluding CNO blending from targeted lines. We additionally compared to the high-resolution optical follow-up analysis of APOGEE stars in clusters from \citet[][]{juliaphd} and O'Connell et al., {\it in prep}, which gives similar results to \citep{apogee_Ce}.  Given the possible uncertainties with cerium, we present the DR17 OCCAM results here, but suggest further work is needed to settle this discrepancy.

\subsection{Comparison of Galactic Abundance Trends}

{In order to evaluate our reported gradients, we compare them against previous studies in this section. For our metallicity gradient comparisons, we use our full sample of open clusters. However, for the individual abundance gradient comparisons in \S \ref{sssec:alpha}-- \ref{sssec:Ce}, we use gradients with a cut in radius at $14$ kpc (Table \ref{tab:gradients}), as the other studies do not have significant clusters beyond $14$ kpc. }

\subsubsection{Galactic Metallicity Gradient}

Our derived metallicity trend --- namely an inner gradient of $-0.073 \pm 0.002$ dex/kpc, outer gradient of $-0.032 \pm 0.002$ dex/kpc, and break at 11.5 kpc for $R_{GC}$ --- shows a very similar inner gradient to that reported in OCCAM-IV (theirs being $-0.068 \pm 0.004$ dex/kpc).  However, our measured outer gradient is significantly steeper than the OCCAM-IV value of $-0.009 \pm 0.011$ dex/kpc, and the knee measured here is farther inwards than theirs (13.9kpc).
These discrepancies are most likely due to poor coverage of clusters at $R_{GC} > 14$ kpcs in the OCCAM-IV sample, as noted in that study. 

The \citet{Netopil21} study finds an overall linear gradient of $-0.058\pm 0.004$ dex/kpc, 
which is only slightly steeper than our measured single linear slope of $-0.056 \pm 0.001$ dex/kpc. They also measure an inner disk ({$R_{GC} < 12$} kpc) gradient of $-0.058 \pm 0.005$ dex/kpc which is significantly shallower than our reported values.   

Similarly, \citet{spina21} measure a linear trend for their sample of open clusters of d[Fe/H]/\rgc $= -0.076 \pm 0.009$ dex/kpc and d[Fe/H]/\rguide $= -0.073 \pm 0.008$ dex/kpc, for clusters between roughly $6 \le R \le 14$ kpc. Both of these slopes are consistent with the present inner gradients of $-0.073 \pm 0.002$ dex/kpc and $-0.074 \pm 0.002$ dex/kpc respectively, however with the additional clusters beyond $\sim 14$ kpc in this sample, we find a much shallower linear gradient in both \rgc and \rguide.

\subsubsection{$\alpha-$Elements -- O, Mg, Si, S, Ca, Ti}\label{sssec:alpha}

Our results for the $\alpha-$elements are largely in agreement with those of OCCAM-IV, with an exception for the gradient in [Ca/Fe] which, in the DR17 sample is significantly flatter ($0.007 \pm 0.002$ dex/kpc) than was reported with the DR16 sample ($0.012 \pm 0.001$ dex/kpc). 
Our gradients for silicon and titanium ($d[{\rm Si/Fe}]/dR_{GC} = +0.001 \pm 0.001$ dex/kpc and $d[{\rm Ti/Fe}]/dR_{GC} = 0.003 \pm 0.003$ dex/kpc) 
are also slightly steeper than those reported in OCCAM-IV 
($d[{\rm Si/Fe}]/dR_{GC} = -0.001 \pm 0.001$ dex/kpc and $d[{\rm Ti/Fe}]/dR_{GC} = 0.000 \pm 0.002$ dex/kpc), but they are nearly the same within the measured uncertainties.

\citet{spina21} find a comparatively steep gradient in [O/Fe] vs \rguide of $0.032 \pm 0.01$ dex/kpc, significantly different from the present measurement of $+0.013 \pm 0.002$ dex/kpc. The steep [O/Fe] vs \rguide gradient stands out from other gradients in $\alpha$ elements in \citet{spina21}, and the reported uncertainty, is larger as well. 
While \citet{casamiquela_2019} reports a {steep gradient for} [Si/Fe] versus \rgc of $0.022 \pm 0.007$, {similar to} the steep [O/Fe] versus \rgc gradient from \citet{spina21}, steep gradients in $\alpha$ elements are not commonly reported. 
Indeed, gradients measured for other $\alpha$ elements (Mg, Si, Ca, and Ti) by \citet{spina21} are nearly flat. This also stands in some contrast to the present work as we consistently measure mildly positive gradients in the same elements. This general trend of mildly positive gradients in $\alpha$ elements is consistent with OCCAM-IV, and previous literature \citep[e.g.,][]{carrera_2011, yong_2012, reddy_16} as discussed in OCCAM-IV.

\subsubsection{Iron-Peak Elements -- V, Cr, Mn, Co, Ni}\label{sssec:ironpeak}

{The gradients we reported in Table \ref{tab:gradients} for nickel, cobalt, manganese, and vanadium are in good agreement with those in OCCAM-IV.  However, the gradient for [Cr/Fe] measured here ($-0.002 \pm 0.005$ dex/kpc) does deviates slightly from that measured in OCCAM-IV ($0.010 \pm 0.004$ dex/kpc). This seems to be due to minute changes in abundances, particularly in those clusters at radii less than $\sim7$ kpc, which affects the gradient and accounts for the discrepancy. } 

{The slopes measured for [Cr/Fe] ($-0.003 \pm 0.004$ dex/kpc) is consistent with the compiled gradient from the Open Cluster Chemical Abundances of the Spanish Observatories survey (OCCASO; \citealt{casamiquela_2019}), $-0.005 \pm 0.003$ dex/kpc. However, the measured slope for [V/Fe], $0.028 \pm 0.011$ dex/kpc, is inconsistent with that measured in OCCASO. This discrepancy can easily be accounted for due to the large scatter present in both gradients.} Both this study and the OCCASO study also measure a very flat gradient for [Ni/Fe], however the final values ($-0.003 \pm 0.002$ dex/kpc in this sample and $0.002 \pm 0.001$ dex/kpc in the OCCASO sample) are just outside of the uncertainties.

{Finally, the slope from this sample for [Mn/Fe] ($-0.011 \pm 0.002$ dex/kpc) is consistent with that reported in \citet{spina21} ($-0.012 \pm 0.004$ dex/kpc). However, the slope reported in this study for the [Ni/Fe] gradient ($-0.004 \pm 0.002$ dex/kpc) is significantly shallower and for the [Co/Fe] gradient ($-0.023 \pm 0.007$ dex/kpc) is significantly steeper than the gradients reported in \citet{spina21} ($-0.022 \pm 0.006$ dex/kpc and $-0.007 \pm 0.007$ dex/kpc, respectively).}

\subsubsection{Odd-Z Elements -- Na, Al, K}\label{sssec:oddZ}

{The gradient calculated for [Na/Fe] and [K/Fe] ($-0.031 \pm 0.006$ dex/kpc and $0.017 \pm 0.003$ dex/kpc ) is consistent with those reported in OCCAM-IV.} However, the slope of the [Al/Fe] gradient, $0.005 \pm 0.003$ dex/kpc, is significantly shallower than the slope reported in OCCAM-IV ($0.018 \pm 0.002$ dex/kpc) and entirely inconsistent with the value reported in \citet{spina21} ($-0.013 \pm 0.007$ dex/kpc). {Additionally, the measured gradient, $d[{\rm Na/Fe}]/dR_{guide} = -0.035 \pm 0.008$ dex/kpc, is inconsistent with that measured in \citet{spina21} ($-0.008 \pm 0.010$ dex/kpc). }However, we note that this sample has a greater number of clusters at larger distances than the \citet{spina21} sample, which flattens the gradient. Additionally, the gradient measured by \citet{spina21} seems to be dominated by a few [Al/Fe] enhanced clusters in the inner galaxy, which we do not see in our sample. {Finally, the OCCAM-IV [Al/Fe] gradient includes a few single-star clusters at low [Al/Fe] which resulted in a steeper measured gradient.}

\subsubsection{The Neutron Capture Element  Ce}\label{sssec:Ce}

Shown in Figure \ref{fig:neutron}, and reported in Table \ref{tab:gradients}, we find a positive {cerium} abundance gradient of $d[{\rm Ce/Fe}]/dR_{guide} = 0.024 \pm 0.006$ dex/kpc and $d[{\rm Ce/Fe}]/dR_{GC} = 0.022 \pm 0.006$ dex/kpc.
Comparing these slopes to the gradients calculated in \citet{apogee_Ce}, 
their value for $d[{\rm Ce/Fe}]/dR_{GC}$ of $0.014 \pm 0.007$ dex/kpc is shallower than the value found here, though much of this discrepancy may be explained by the measurement differences as described in \S \ref{sec:other_surveys}. 

\subsection{Evolution of Galactic Abundance Gradients}
\subsubsection{Iron}\label{sec:fe_evo}

One of the key goals of the OCCAM project is to explore the evolution of abundance gradients in the Milky Way.
To this end, we find significant evolution in the $d[{\rm Fe/H}]/dR$ gradients as presented in \S \ref{sec:FeH}.
This same trend has also been shown in OCCAM-IV, \citet{spina21},  \citet{apogee_Ce}, \citet{Netopil21} and \citet{Zhang21}. 

\begin{figure}
    \epsscale{1.2}
 	\plotone{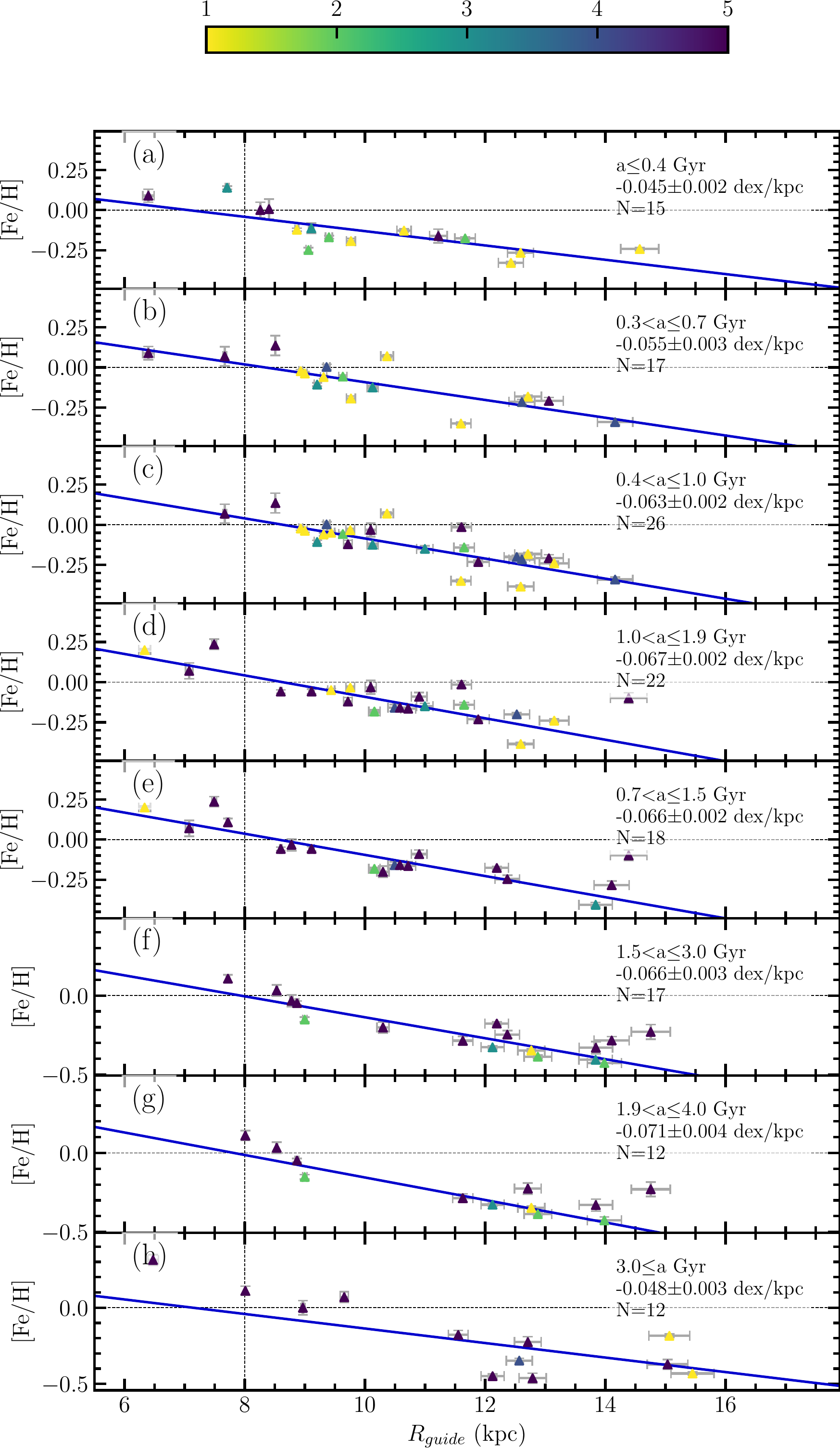}
 	\caption{ \small The age slopes measured if the age bins used in \citep{Netopil21} are adopted. }
 	\label{fig:netopil_comp}
\end{figure}

\citet{Netopil21} explored the evolution of the [Fe/H] gradient by compiling a sample of 136 open clusters from various studies, including 75 clusters with data from APOGEE DR16, 70 of which are in common with this sample. The details of the compilation are recorded in both \citet{netopil_16} and \citet{Netopil21}. The latter used this sample to investigate radial migration {in open clusters} and also measure the age-metallicity gradient with eight overlapping age bins. These age bins span from the youngest clusters (age $ < 0.4$ Gyrs) to clusters with age $\ge 5.2$ Gyrs.

To better compare our results to that of \citet{Netopil21}, we divided our sample into their age bins (Fig. \ref{fig:netopil_comp}); however, to populate the oldest age bin with more than 10 clusters, we modified the oldest age bin from the Netopil et al. limits $3.0 \ge$ age $\ge 5.2$ Gyrs to instead include all clusters with age $\ge 3.0$ Gyr. 

Comparing to Table 6 in \citet{Netopil21}, the gradients we measure in Figure \ref{fig:netopil_comp} are in good agreement for nearly every age bin, with measurements in {6} of the 8 samples agreeing well within the reported uncertainties. However, in the {first} age bin, the discrepancy {between the two gradients} is $\sim 0.004$ dex/kpc; and for the final age bin the discrepancy is $\sim 0.008$ dex/kpc. {We note that the final age bin is largely affected by two relatively metal poor clusters at \rguide $\simeq 12$ kpc, NGC 2243 and Trumpler 5.}

\begin{figure*}
    \epsscale{1.15}
 	\plotone{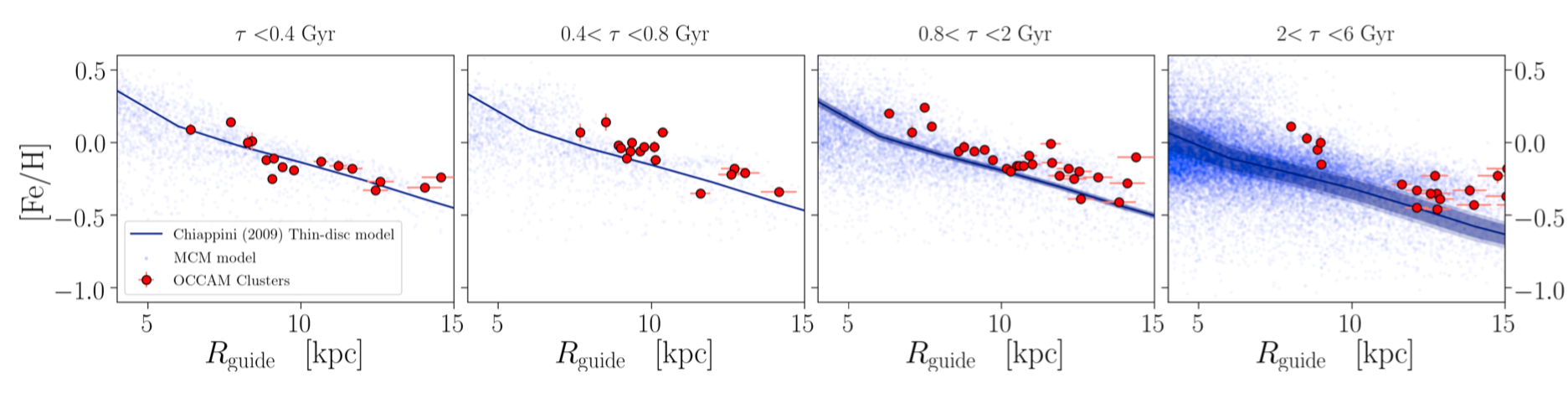}
 	\plotone{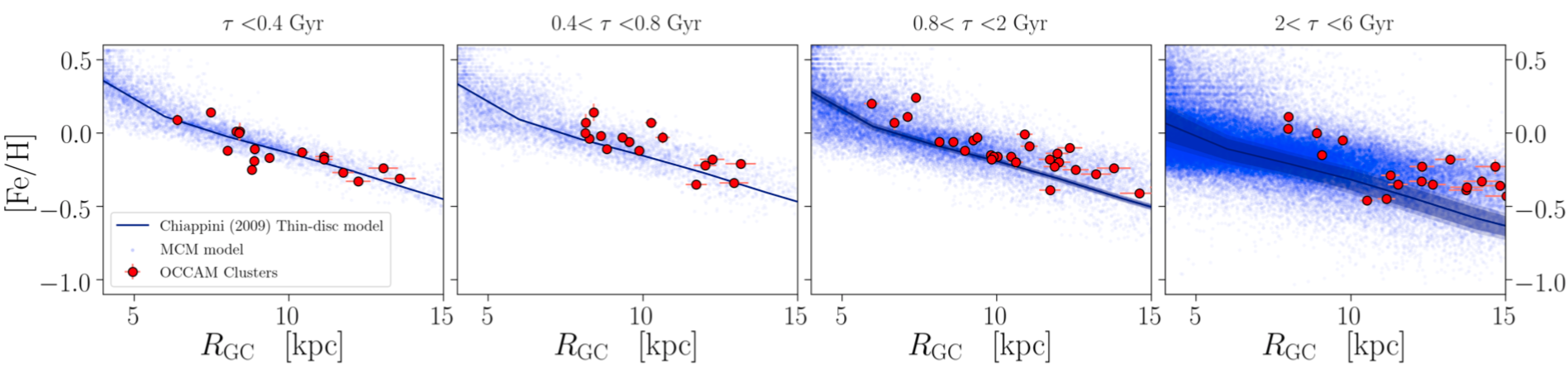}
 	\caption{ \small The open cluster sample from this study (red dots) overlaid on the models of \citep{chiappini_2009} (the blue line) and \citep{minchev_1,minchev_2} (blue dots). The plots are split into the same age bins used in Figure \ref{fig:comprad}.}
 	\label{fig:Model_comp}
\end{figure*}

Finally, we compare our [Fe/H] evolution results to the thin-disk chemical evolution model of \citep{chiappini_2009} and the chemo-dynamical simulation of \citep[][MCM in Figure \ref{fig:Model_comp}]{minchev_1,minchev_2}, using the same age bins as Figure \ref{fig:comprad}. 
In the youngest three age bins we notice good agreement for both \rgc and \rguide trends. The third age bin also shows decent agreement, though there may be a slight offset between the \rguide cluster sample and the model results.
In the final age bin, both the \rgc and \rguide cluster sample have a noticeable offset from the models. This could potentially suggest either a real effect that would require a change to the models or that the older open clusters are possibly a biased sample due to which clusters survive to older ages and/or that these old clusters may have undergone migration and migrated outward during their lifetimes (whereas clusters that moved inward in the Milky Way are more likely to be disrupted).

\subsubsection{[X/Fe]}\label{elem_evo}

The evolution of abundance gradients for elements besides iron were explored in \S \ref{all_elems}.
A similar analysis in OCCAM-IV indicated no convincing trends in $\alpha$ elements with time. As in OCCAM-IV, it could be argued that there is a slight trend for [Mg/Fe], with older clusters perhaps showing generally steeper slopes than the younger clusters, but also as in OCCAM-IV the changes between samples are roughly as significant as the uncertainties.

There is very little evolution found for [Cr/Fe] and [Ni/Fe], which is consistent with the OCCAM-IV results.  As is the case here, OCCAM-IV found significant uncertainty in the [Cr/Fe] measurements, but the [Ni/Fe] gradients were fairly well determined. 
OCCAM-IV found a significant trend for [Mn/Fe] with the gradient becoming more negative for younger cluster populations. With the new DR17 data, this trend is no longer present; indeed the new APOGEE results seem to indicate that the younger samples have less negative gradients.

{Finally, while we do seem to see an increasingly negative trend in d[Na/Fe]/d\rguide as clusters get younger, the uncertainties in the gradients are large. This is roughly consistent with the slopes calculated in OCCAM-IV, though they found a flatter trend with a significantly steeper slope in the oldest age bin that we do not see here. }

\section{Conclusions }

We present the final APOGEE-2 DR17 OCCAM sample, which consists of 150 open clusters, and 94 that we designate as ``high quality''. 
To gain insights into the chemical enrichment history of the Milky Way, we use the high quality sample to measure Galactic abundance gradients in 16 chemical elements and investigate their evolution over four age bins.
With clusters spanning roughly 6.0 to 18 kpc, we measure a two-function Galactic radial metallicity trend, with $-0.073 \pm 0.002$ dex/kpc for the inner slope, $-0.032 \pm 0.002$ dex/kpc for the outer slope, and a knee located at 11.5 kpc. 
In order to account for blurring effects in the clusters orbits, we also calculate the guiding center radii, $R_{guide}$, of each cluster. By using \rguide as the independent variable, we find an inner slope of $-0.074 \pm 0.002$ dex/kpc, an outer slope of $-0.023 \pm 0.003$ dex/kpc, and a knee at 12.2 kpc. 

The Galactic radial gradients for the 15 elements measured in this survey are in good agreement with other recent studies \citep[e.g.,][]{reddy_16,casamiquela_2019,occam_p4,spina21}. In this work, we find significant ($3\sigma$ or greater) trends in 9 of the 15 elements, including four of the $\alpha-$elements (O, Mg, S, Ca), all of the odd-Z elements (Na, Al, K), and cerium. We don't find significant gradients in the iron-peak elements, except manganese.  

We explore the variation in the trends for all elements throughout time, by splitting the open cluster sample into four age bin.  We find no significant evolution compared with solar ratios, besides two elements (V and Na) which have large uncertainty in their measurements. {This lack of age variation in the gradients points to well-mixed enrichment through the age range covered (10 Myr -- 9 Gyr), which implies that chemical tagging distinct {\it age} populations may be difficult with these elements, but could be improved with the inclusion of C and N \citep[e.g.,][]{casali19, spoo_22} for distinct stellar evolutionary phases.}

We compare this DR17-based sample to OCCAM-IV and the GALAH sample from \citet{spina21} and find no significant differences between the abundances in either case. Additionally, we compare against the cerium abundances derived in \citet{apogee_Ce} and find an abundance correlated offset for sub-solar cerium abundances between the BACCHUS analyses and the DR17 ASPCAP-derived values.

We find general agreement in the first three age-bins when we compare to the chemo-dynamical models of \citet{chiappini_2009} and \citet{minchev_1, minchev_2}, however in the final age bin we do find an offset between the cluster sample and the models. This could be explained by either an offset in the models or, possibly, by a potential survivor bias in the older open cluster sample.

Also, we note that APOGEE DR17 is able to measure Galactic trends for many of the CHNOPS elements, e.g., C,N,O, and S,
which are important in the astrobiological study of the Galactic habitable zone. 
In this work, we present the gradients for oxygen and sulfur. The gradients for carbon and nitrogen are not presented here due to stellar evolutionary effects that change stellar surface chemistry due to the dredge up; however, these elements and their correlations with age are explored in \citet{spoo_22}.

\begin{acknowledgements}

We would like to greatly thank Friedrich Anders for helping with the creation of figures used for comparison to the Chiappini model.  
\end{acknowledgements}

NM, JD, PMF, TS, and AER acknowledge support for this research from the National Science Foundation (AST-1715662). PMF and APW acknowledge this work was performed at the Aspen Center for Physics, which is supported by National Science Foundation grant PHY-1607611.
DAGH acknowledges support from the State Research Agency (AEI) of the Spanish Ministry of Science, Innovation and Universities (MCIU) and the European Regional Development Fund (FEDER) under grant AYA2017-88254-P.
DG gratefully acknowledge support from the Chilean Centro de Excelencia en Astrof\'isica
and financial support from the Dirección de Investigaci\'on y Desarrollo de
la Universidad de La Serena through the Programa de Incentivo a la Investigaci\'on de
Académicos (PIA-DIDULS).
DM is supported by ANID BASAL project FB210003.
HJ acknowledges support from the Crafoord Foundation, Stiftelsen Olle Engkvist Byggm\"astare, and Ruth och Nils-Erik Stenb\"acks stiftelse.
ARL acknowledges financial support provided in Chile by Comisi\'on Nacional de Investigaci\'on Cient\'ifica y Tecnol\'ogica (CONICYT) through the FONDECYT project 1170476 and by the QUIMAL project 130001

Funding for the Sloan Digital Sky 
Survey IV has been provided by the 
Alfred P. Sloan Foundation, the U.S. 
Department of Energy Office of 
Science, and the Participating 
Institutions. 

SDSS-IV acknowledges support and 
resources from the Center for High 
Performance Computing  at the 
University of Utah. The SDSS 
website is www.sdss.org.

SDSS-IV is managed by the 
Astrophysical Research Consortium 
for the Participating Institutions 
of the SDSS Collaboration including 
the Brazilian Participation Group, 
the Carnegie Institution for Science, 
Carnegie Mellon University, Center for 
Astrophysics | Harvard \& 
Smithsonian, the Chilean Participation 
Group, the French Participation Group, 
Instituto de Astrof\'isica de 
Canarias, The Johns Hopkins 
University, Kavli Institute for the 
Physics and Mathematics of the 
Universe (IPMU) / University of 
Tokyo, the Korean Participation Group, 
Lawrence Berkeley National Laboratory, 
Leibniz Institut f\"ur Astrophysik 
Potsdam (AIP),  Max-Planck-Institut 
f\"ur Astronomie (MPIA Heidelberg), 
Max-Planck-Institut f\"ur 
Astrophysik (MPA Garching), 
Max-Planck-Institut f\"ur 
Extraterrestrische Physik (MPE), 
National Astronomical Observatories of 
China, New Mexico State University, 
New York University, University of 
Notre Dame, Observat\'ario 
Nacional / MCTI, The Ohio State 
University, Pennsylvania State 
University, Shanghai 
Astronomical Observatory, United 
Kingdom Participation Group, 
Universidad Nacional Aut\'onoma 
de M\'exico, University of Arizona, 
University of Colorado Boulder, 
University of Oxford, University of 
Portsmouth, University of Utah, 
University of Virginia, University 
of Washington, University of 
Wisconsin, Vanderbilt University, 
and Yale University.

This work has made use of data from the European Space Agency (ESA) mission {\it Gaia} (\url{https://www.cosmos.esa.int/gaia}), processed by the {\it Gaia} Data Processing and Analysis Consortium (DPAC, \url{https://www.cosmos.esa.int/web/gaia/dpac/consortium}). Funding for the DPAC has been provided by national institutions, in particular the institutions participating in the {\it Gaia} Multilateral Agreement.

This research made use of Astropy, a community-developed core Python package for Astronomy (Astropy Collaboration, 2018).


\facilities{Du Pont (APOGEE), Sloan (APOGEE), Spitzer, WISE, 2MASS, Gaia}
\software{\href{http://www.astropy.org/}{Astropy}} 

\bibliography{Myers}{}
\bibliographystyle{aasjournal}



\end{document}